\documentclass{cta-author}
\usepackage{graphicx}
\usepackage{picinpar}
\usepackage{amsmath}
\usepackage{url}
\usepackage{flushend}
\usepackage[latin1]{inputenc}
\usepackage{colortbl}
\usepackage{soul}
\usepackage{multirow}
\usepackage{pifont}
\usepackage{color}
\usepackage{alltt}
\usepackage{enumerate}
\usepackage{siunitx}
\usepackage{breakurl}
\usepackage{epstopdf}
\usepackage{pbox}
\usepackage{subfigure}
\usepackage{comment}
\usepackage{hyperref}
\renewcommand{\figurename}{Fig.}

\begin{document}

\title{Sensorless Real-Time Reduced Order Model Based Adaptive Maximum Power Tracking Pitch Controller for Grid Connected Wind Turbines}

\author{\au{Abilash Thakallapelli$^{1}$},
\au{Sudipta Ghosh$^{1}$},
\au{Sukumar Kamalasadan$^{1\corr}$}}

\address{\add{1}{Electrical and Computer Engineering, UNC Charlotte, Charlotte, USA}
\email{skamalas@uncc.edu}}

\begin{abstract}
This paper presents a sensor-less maximum power tracking (MPT) pitch controller for grid connected Wind Turbine (WT). The main advantage of the proposed architecture is that the approach ensures smooth operation and thus minimizes the mechanical stress and damage on the WT during high wind speed and grid transient conditions. Simultaneously, it also: a) reduces transients in Point of Common Coupling (PCC) bus voltage, b) reduces rotor speed oscillations, and c) controls the output power of the wind turbine without exceeding its thermal limits. The approach can work without wind speed measurements.  In order to consider the effect of grid variations at the PCC, the affected area in the grid is modeled as a study area (area of interest), and remaining area (external area) is modeled as frequency dependent reduced order model (FDROM). The reduced order model (ROM) is then used to estimate the reference speed.  The proposed controller is designed using the error between actual speed of the generator and the reference speed, to ensure smooth operation and limit the speed and aerodynamic power at the rated values. The architecture is evaluated using wind farm integrated Kundur's two-area and IEEE-39 bus test systems using real-time digital simulator (RTDS).

\end{abstract}

\maketitle

\section{Introduction}
In recent years, the move towards eliminating fossil fuel dependency and embracing sustainable energy based power generation has increased interest in integrating renewable energy sources (RES) into the power grid. In 2016, WTG provided almost 6\% of U.S. electricity generation (about 37\% of electricity generation from RES) \cite{ref1}. However, WTGs operate under varying wind conditions and depends on time and geographical location, which may be above or below rated values, thus varying their output power. During high wind speed conditions, the controller should limit the speed of the generator not crossing the rated value by limiting the rotation rate of the rotor, since pitch system contributes to 21.3\% of the overall failure rate of wind turbines \cite{ref2a}. This can be achieved by controlling the blade pitch angle \cite{ref2}. However, in practical systems, WTG operations are also influenced by the dynamics of the entire power grid. Thus, the design of WTG controllers should take into consideration of grid dynamics. Wind speed conditions are generally measured using anemometers, failure of which can cause deterioration in tracking performance. This should be addressed in the controller design as well \cite{ref4}.

Grid level interactions of the wind farms/turbines are generally controlled considering a constant voltage at the PCC even though electro-mechanical dynamics are included in such simulations. This ignores response of the wind farms with the electromagnetic transients in the grid.  The effect is on the mechanical fatigue that happens on the wind generators. If one should design a controller considering grid dynamics, detailed Electro-Magnetic Transient (EMT) based grid models with dynamic models of WTG including DFIG are required.  However, detail modeling of large scale power grid is impractical due to computational complexity \cite{ref3}. In \cite{ref3,ref5}, to reduce computational burden several model order reduction techniques based on linearized models have been developed, but these models are effective only during low-frequency oscillations. A possible method to reduce computational burden while retaining accuracy is to model part of WTG integrated grid (study area) in detail and the remainder of the network (external area) \cite{ref6} as an equivalent. For this, the external area is modeled as a combination of low frequency (Transient Stability Assessment -TSA type) and high frequency (FDNE type) equivalents. In TSA type, the network is formulated as an admittance matrix at the fundamental frequency, and the generators are aggregated and modeled in detail such that low-frequency electromechanical oscillations are preserved, whereas the high-frequency oscillations are preserved by FDNE.

In the literature, several WTG pitch control strategies for limiting the aerodynamic power and generator speed are proposed. An individual pitch control scheme with a proportional-integral (PI) controller with two resonant compensators is proposed in \cite{ref7}. However, the PI controllers are designed based on a specific operating point. A pitch angle controller based on fuzzy logic is proposed in \cite{ref8}, in which generator output power and speed are used as input to the controller. However, determining exact fuzzy rules and membership functions for a dynamically changing conditions are considered. In \cite{ref9}, a fuzzy predictive algorithm coupled with conventional PI controllers is proposed for wind-turbine collective-pitch control. In \cite{ref10}, a method of nonlinear PI control for variable pitch wind turbine is proposed. The non-linearities and disturbances are evaluated and compensated using extended order state and perturbation observer. However, this method uses only one set of PI parameters for various speeds. Ref. \cite{ref11} investigated determining the pitch angle when wind speed exceeds rated value using particle swarm optimization (PSO) and \cite{ref12} proposed a method for blade pitch angle control using PID control. 

In this paper, a novel sensor-less method for smoothly controlling the transients of WTG during high wind speed is introduced. The architecture uses an online dynamic network model of the power grid  that is computationally tractable, to calculate reference speed for tracking. Then an adaptive controller is desgined for smooth tracking and limiting the mechanical stress on the turbine. The control variables used are the algebraic error between the calculated reference speed and actual generator speed. For controller adaptation, a model identification method based on Recursive Least Square (RLS) method is also designed \cite{ref13}. RLS identification is performed online to estimate the transfer function with the difference between the reference and actual speed as the process output and the controlling signal as the process input. Then using the identified transfer function, the controller gains of the controller are calculated online. If there is a change in operating point, the controller auto-tunes as the transfer function is identified every sample time. This auto-tuning feature allows the proposed controller to provide an efficient way for adjusting the pitch angle during changing system operating conditions, as opposed to the conventional PI controller where gains are constant irrespective of the system conditions.
\subsection{Contributions}

The advantages of the proposed architecture are it,

\begin{itemize}
	\item auto-tunes based on the wind speed and grid conditions and thus can higher precision.
	\item can be implemented in practical systems as the online grid models are computationally tractable. 
	\item provides dynamic control capabilities as opposed to conventional controllers.
	\item can eliminate the requirement of anemometer.
	\item reduces mechanical stress on the turbine, voltage transients and speed oscillations. 
\end {itemize}	
\subsection{Paper Organization}	
	
	The rest of the paper is organized as follows. In section II the wind turbine and generator modeling are discussed. In section III, frequency dependent reduced-order modeling of the large power grid is discussed. Section IV discusses the proposed adaptive pitch controller and example case study.  Section V discusses the implementation of TSA/FDNE and the proposed control architecture in Real-time Digital Simulator. Section VI discusses the real-time simulation results and section VII concludes the paper.
\section{Wind Turbine and Generator Modeling}

The variable speed WTGs are more frequently involved in providing grid reliability as they are more controllable, provides reactive power support and harvests optimum energy over a wide wind speed range \cite{ref14}-\cite{ref16}. In this paper, a two-mass variable speed model of WTG is designed and scaled up to represent 200 MW of rated power at the VSC interface transformer for modeling purposes.
\subsection{The Wind Turbine}
The mechanical power output ($P_m$) of the turbine in kW \cite{ref17} can be represented as 
\begin{equation}
\label{eqn1}
P_m = C_p(\lambda,\beta)\frac{\rho A}{2}v^3_{wind}
\end{equation}
where $C_p(\lambda,\beta)$  is the coefficient of performance of the turbine which can be determined from the $C_p$vs$\lambda$ curve for different blade pitch angle $(\beta)$, $\lambda$  is the tip speed, $\rho$  is the density of air in  $kg/m^3$,   $A$ is the area swept by the turbine blades in $m^2$, and $v_{wind}$  is the velocity of the wind in m/s. From this, $\lambda$  can be represented as \cite{ref16a,ref16b}
\begin{equation}
\label{eqn2}
\lambda = \frac{R\omega_t}{v_{wind}}
\end{equation}
where $R$ and $\omega_t$, are the radius of the turbine $(m)$ and the rotational speed of the turbine $(rad/s)$ respectively.
\subsection{The Coefficient of Performance}
The turbine coefficient of performance describes the power extraction efficiency of the WT and is generally less than 0.5. This can be represented as \cite{ref18}
\begin{equation}
\label{eqn3}
C_p(\lambda,\beta)=c_1\left[\frac{c_2}{\lambda_i}-c_3\beta-c_4\right]e^{-\frac{c_5}{\lambda_i}}+c_6\lambda
\end{equation}
where $$\frac{1}{\lambda_i}=\frac{1}{\lambda+0.08\beta}-\frac{0.035}{\beta^3+1}$$

\begin{figure}
\centering     
\subfigure[$C_p$ vs $\lambda$ curve.]{\label{fig1a}\includegraphics[width=3.5in]{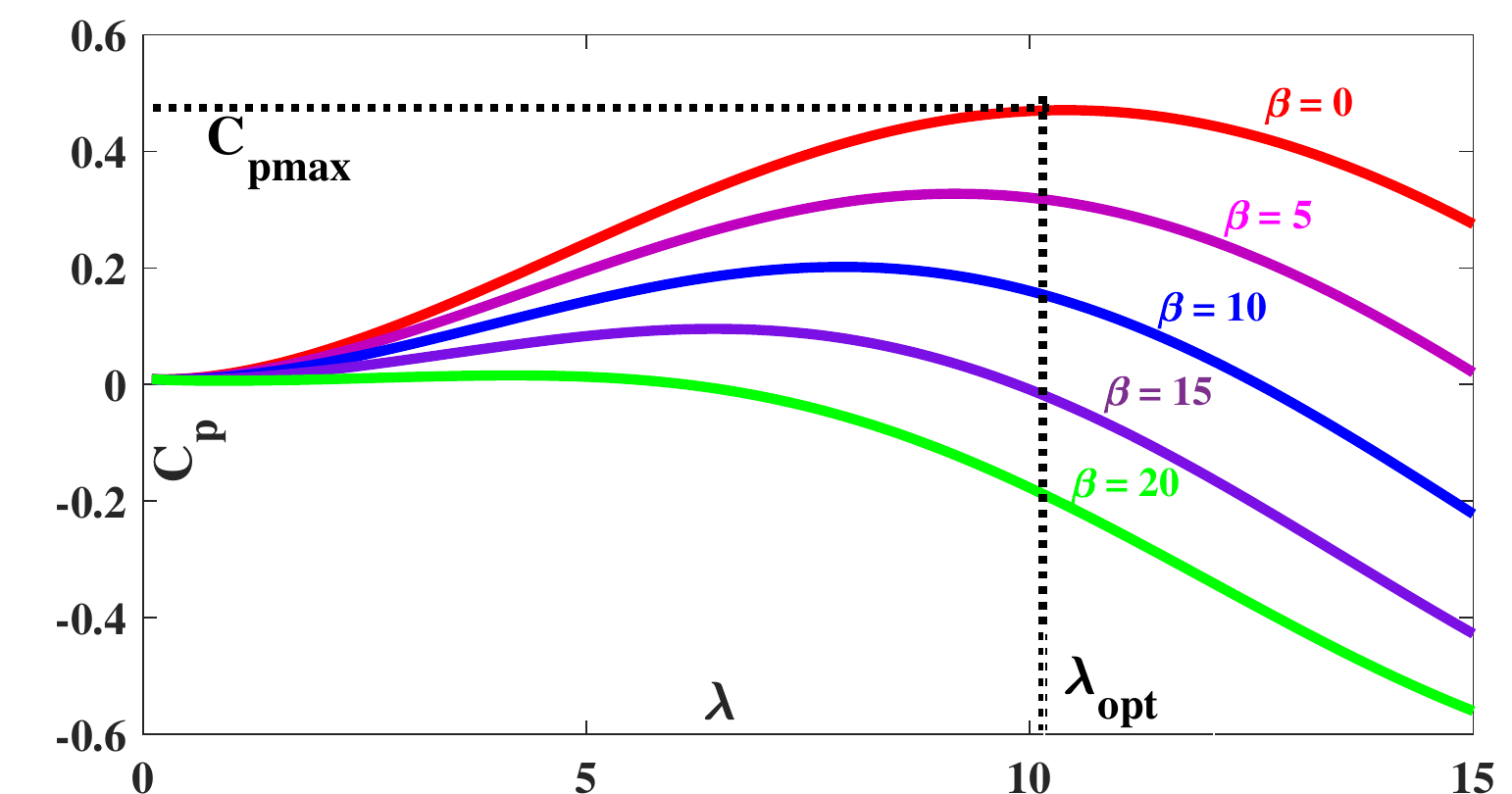}}
\subfigure[Power output from the turbine for various wind speeds]{\label{fig1b}\includegraphics[width=3.5in]{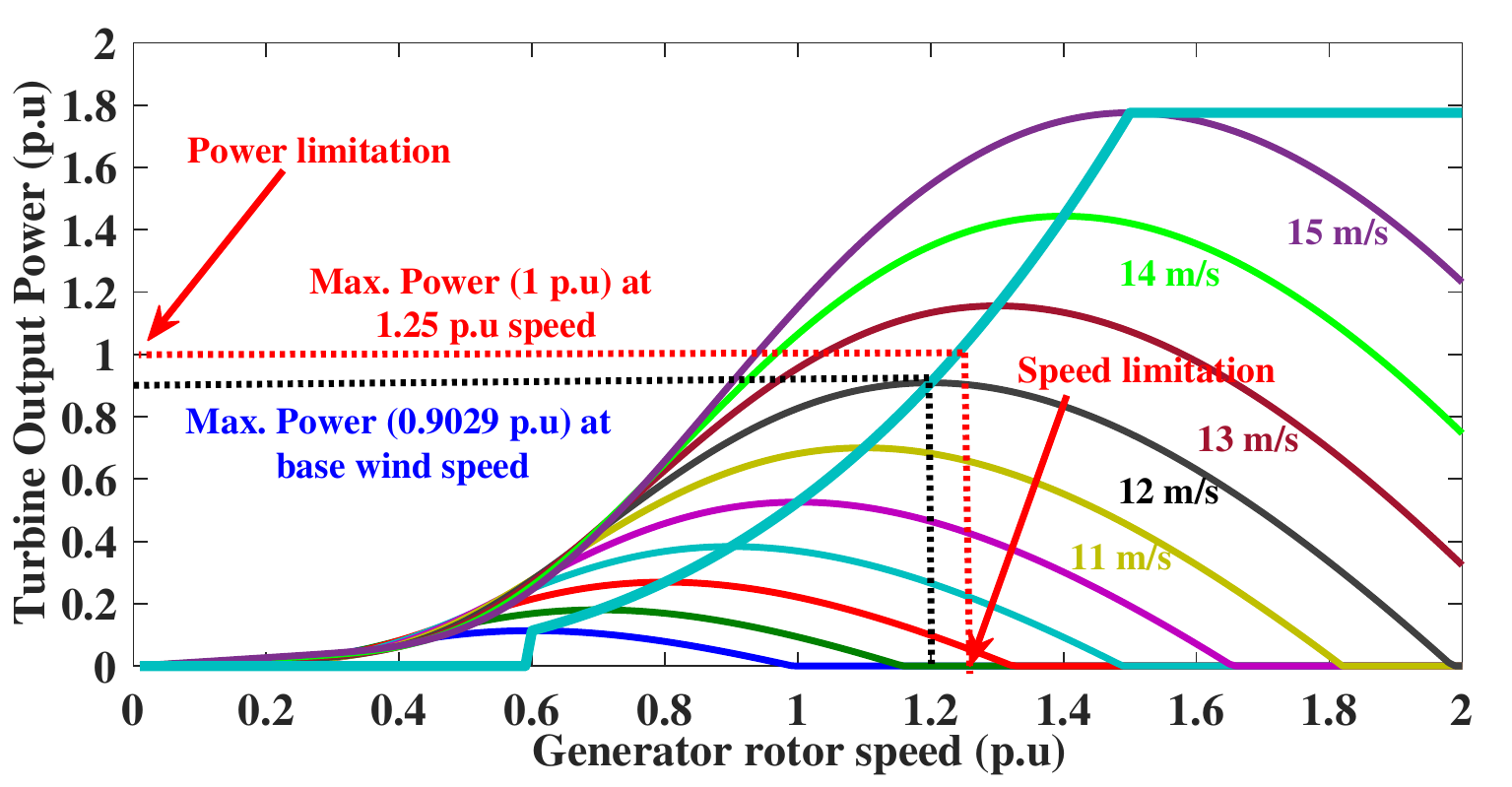}}
\caption{Wind turbine characteristics.}
\end{figure}

For the proposed design,  $c_1=0.5176$,  $c_2=116$,  $c_3=0.4$,  $c_4=5$,  $c_5=21$ and  $c_6=0.0068$. 
The value of tip speed ratio $\lambda$  is constant for all maximum power points. The maximum value for power coefficient $C_p$ for a particular wind turbine can be obtained from $C_p$vs$\lambda$  curve for different values of $\beta$. For the wind turbine selected for this work, the optimum value and the maximum value of  $\lambda$ are 10.4 and 0.48 respectively at $\beta=0^o$.  A characteristic plot of  $C_p$ vs $\lambda$ for the proposed turbine based on \eqref{eqn3} is as shown in Fig. \ref{fig1a}. From Fig. \ref{fig1a}, it can be observed that as $\beta$  increases, $\lambda$ decreases due to a decrease of turbine speed, and simultaneously $C_p$  becomes less. This feature is used in pitch angle control to limit the speed of the rotor for wind speeds greater than the rated value. Fig. \ref{fig1b} shows the turbine output power (p.u) vs rotor speed (p.u) for various wind speeds.
\subsection{Wind Generator}
In the proposed study, type III DFIG with conventional vector control based Grid Side and Rotor Side Controllers is considered. The detailed nonlinear model of DFIG is developed in RSCAD\textsuperscript{TM}. Modeling details of DFIG are discussed in several previous works \cite{ref15}, \cite{ref19,ref20}.
\subsection{Maximum Power Point Tracking (MPPT)}
At any speed, from \eqref{eqn1}
\begin{equation}
\label{eqn4}
P_m = k_pc_pv^3_w = k_pc_p\left(\frac{\omega_r}{r_{gear}\lambda}\right)^3
\end{equation}

\begin{equation}
\label{eqn5}
\omega_r = r_{gear}\lambda\left(\frac{P_m}{k_pc_p}\right)^\frac{1}{3}
\end{equation}

where $\omega_t$ and $\omega_r$ [p.u] are the angular speed of the turbine and rotor respectively, and $P_m$ is the turbine mechanical power in [p.u]. The scaling factor  $k_p\left(=\frac{\rho Ac_{pmax}\times v_{\omega BASE}}{2P_{BASE}}\right)$ indicates maximum output power at base wind speed.  The angular speed of the turbine,   $\omega_t$ [p.u], is related to the generator rotor speed by the gear ratio, ($r_{gear} = 1.2$), i.e. $\omega_t=\frac{\omega_r}{r_{gear}}$.

\section{Frequency-Dependent Reduced Order Modeling of Power Grid}
Large power systems can be modeled as an equivalent to reduce complexity and computational burden while preserving the high and low-frequency behavior of the system under consideration. To this effect, the proposed frequency dependent reduced-order power system models the area of interest (study) area in detail and the remaining part as a combination of FDNE and coherency based TSA equivalent. First, FDNE is formulated based on online RLS identification, by short and open circuiting all voltage and current sources respectively and energizing the external area with constant voltage and varying frequency.  The FDNE is represented as a discrete transfer function and rearranged as shown in \eqref{eqn7}.
\begin{equation}
\label{eqn7}
\begin{aligned}
I_b(k) = -a_1I_b(k-1)-a_2I_b(k-2)\dots-a_nI_b(k-n)\\
+b_1V_b(k-1)+b_2V_b(k-2)\dots+b_nV_b(k-n)
\end{aligned}
\end{equation}
where  $I_b$ and $V_b$  are the boundary bus current and voltages respectively, $k$ is the current sample and $n$  is the order. 

\begin{figure}[!t]
\centering
\includegraphics[width=3.5in,height=1.9in]{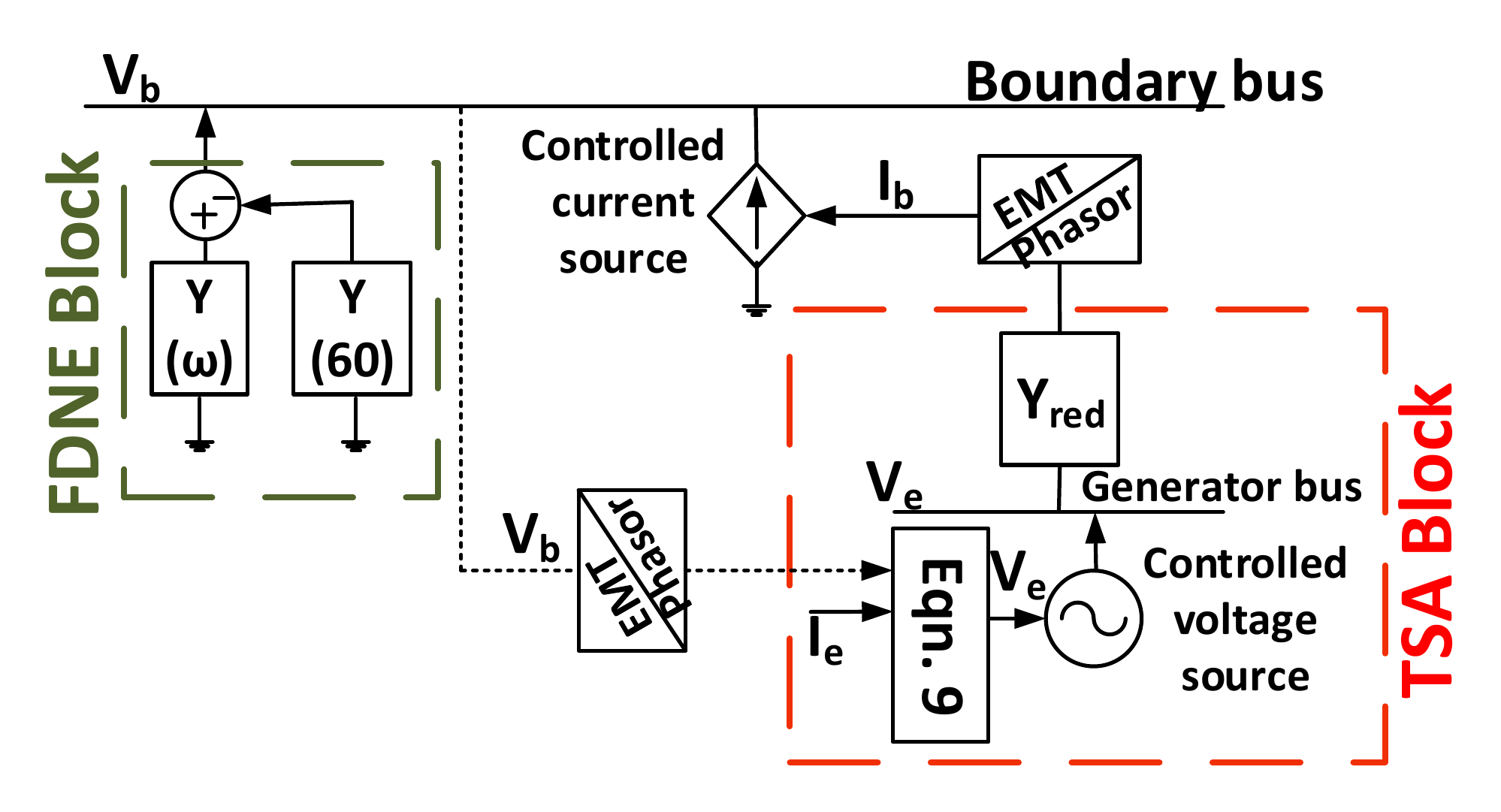}
\caption{FDNE and TSA block diagram for a power network.}
\label{fig3}
\end{figure}

For designing the TSA equivalent and to further reduce the complexity and computational burden, all generating units and nodes in external area are aggregated using coherency based inertial aggregation \cite{ref21,ref22} and the admittance matrix $(Y)$ of external area is reduced to  $Y_{red} (2\times2)$ matrix by Kron node reduction method represented as follows:
\begin{eqnarray}
   \left[\begin{array}{c}
    I_b \\
    I_e
   \end{array}\right]=
    \left[\begin{array}{cc}
    Y_{bb} & Y_{be} \\
    Y_{eb} & Y_{ee}
   \end{array}\right] \left[\begin{array}{c}
    V_b \\
    V_e
   \end{array}\right]
\label{eqn8}
\end{eqnarray}
where $I_e$ and $V_e$  are the generator bus current and voltage respectively. The generator bus voltage is calculated as shown in \eqref{eqn9} and generator bus is energized with $V_e$  as shown in Fig. \ref{fig3}.
\begin{equation}
\label{eqn9}
V_e=\left[I_e-Y_{eb}V_b\right]Y^{-1}_{ee}
\end{equation}
Finally, $I_b$ is calculated as shown in \eqref{eqn10} and injected into the boundary bus.
\begin{equation}
\label{eqn10}
I_b=Y_{bb}V_b+Y_{be}V_e
\end{equation}
The advantage of this method is that the reduced power system model behaves as the original system and can replace the original system for further dynamic assessment of renewable energy sources. Further details regarding reduced order modeling are discussed in \cite{ref23,ref24}.
\section{Proposed Adaptive Pitch Controller}
The proposed adaptive pitch controller involves two steps: 1) Recursive Least Square Identification and 2) Calculating gains of the controller.
\subsection{Recursive Least Square Identification}
The RLS identification with the process input $u(k)$  and the process output $y(k)$  is performed dynamically at every sample $k$. The $n^{th}$ order process of the model in z-domain can be represented as \begin{equation}
\label{eqn11}
\frac{y(k)}{u(k)}=\frac{b_1z^{-1}+b_2z^{-2}+\dots+b_nz^{-n}}{1+a_1z^{-1}+a_2z^{-2}+\dots+a_nz^{-n}}
\end{equation}
where $a's$ and $b's$ are the denominator and numerator coefficients of the transfer function respectively. 
Let $N$  be the observation window length, then \eqref{eqn11} can be rewritten as
\begin{eqnarray}
   \left[\begin{array}{c}
    y(k) \\
    y(k-1) \\
    . \\
    . \\
    . \\
     y(k-N+1)
   \end{array}\right]_{N\times1}=\left[X_{N\times2n}\right]
   \left[\begin{array}{c}
    a_1 \\
    . \\
    . \\
    a_n \\
    b_1 \\
    . \\
    . \\
    b_n
   \end{array}\right]_{2n\times1}
\label{eqn12}
\end{eqnarray}
Equation \eqref{eqn12} can be represented in the generic form as follows
\begin{equation}
\label{eqn13}
\Phi_{model(N\times1)}=X_{N\times2n}\Theta_{2n\times1}
\end{equation}
where $\Phi$ is a matrix of past and current outputs $(y)$, $X$ is a matrix of past inputs and outputs and, $\Theta$ is a  matrix of the numerator and denominator coefficients of the transfer function. If the identified model is different form measurements, then
\begin{equation}
\label{eqn14}
\epsilon=\Phi_{measured}-\Phi_{model}
\end{equation}
where $\epsilon$ is the error between the measurements from the system (subscript measured) and the identified model (subscript model) for which criteria $J$  can define as
\begin{equation}
\label{eqn15}
J=\epsilon^t\epsilon
\end{equation}
By letting $dJ/d\Theta=0$, we get
\begin{equation}
\label{eqn16}
\Theta=\left[X^tX\right]^{-1}X^t\Phi_{measured}
\end{equation}
From \eqref{eqn16}, to identify the coefficients of the transfer function the inverse of the state matrix should be computed. If the size of the state matrix is large, inverting a large matrix will slow down the process and sometimes may be even not achievable. To overcome this issue, a recursive least squares technique is used. RLS is a computational algorithm that recursively finds the coefficients of the model and eliminates the matrix inversion. 
Let $S=X^tX$ then \eqref{eqn16} can be written as 
\begin{equation}
\label{eqn17}
\Theta=S^{-1}X^t\Phi
\end{equation}
where $\Phi=\Phi_{measured}$
\begin{eqnarray}
   \Theta(k)=S^{-1}\left[x(k)X^t(k-1)\right]
    \left[\begin{array}{c}
    \Phi(k) \\
    \Phi(k-1) 
   \end{array}\right]
\label{eqn18}
\end{eqnarray}
\begin{equation}
\label{eqn19}
\Theta(k)=S^{-1}\left[x(k)\Phi(k)+X^t(k-1)\Phi(k-1)\right]
\end{equation}
Using \eqref{eqn13}
\begin{equation}
\label{eqn20}
\begin{aligned}
\Theta(k)=S^{-1}\left[x(k)\Phi(k)+X^t(k-1)X(k-1)\Theta(k-1)\right]
\end{aligned}
\end{equation}
\begin{equation}
\label{eqn21}
\Theta(k)=S^{-1}\left[x(k)\Phi(k)+S(k-1)\Theta(k-1)\right]
\end{equation}
\begin{equation}
\label{eqn22}
S(k)=S(k-1)+x(k)x'(k)
\end{equation}
Substituting \eqref{eqn22} in \eqref{eqn21}
\begin{equation}
\label{eqn23}
\Theta(k)=S^{-1}\left[x(k)\Phi(k)+\{S(k)-x(k)x'(k)\}\Theta(k-1)\right]
\end{equation}
\begin{equation}
\label{eqn24}
\begin{aligned}
\Theta(k)=\Theta(k-1)+[S(k-1)+x(k)x'(k)]^{-1}x(k)\\
[\Phi(k)-x'(k)\Theta(k-1)]
\end{aligned}
\end{equation}
Let $P(k)=S^{-1}(k),$ and by matrix inversion lemma $P(k)$ can be represented as 
\begin{equation}
\begin{aligned}
\label{eqn25}
P(k)=P(k-1)\left[I-\frac{x(k)x'(k)P(k-1)}{1+x'(k)P(k-1)x(k)}\right]
\end{aligned}
\end{equation}
Substituting \eqref{eqn22} in \eqref{eqn21} and letting
\begin{equation}
\begin{aligned}
\label{eqn26}
K(k)=\frac{P(k-1)x(k)}{1+x'(k)P(k-1)x(k)}
\end{aligned}
\end{equation}
where $P(k)$ can be written as
\begin{equation}
\begin{aligned}
\label{eqn27}
P(k)=\left[I-K(k)x'(k)\right]P(k-1)
\end{aligned}
\end{equation}
Therefore, \eqref{eqn24} can be represented as

\begin{equation}
\begin{aligned}
\label{eqn28}
\Theta(k)=\Theta(k-1)+K(k)\left[\Phi(k)-x'(k)\Theta(k-1)\right]
\end{aligned}
\end{equation}
With weighted least square, \eqref{eqn26} and \eqref{eqn27} can be presented as

\begin{equation}
\begin{aligned}
\label{eqn29}
K(k)=\frac{P(k-1)x(k)}{\gamma+x'(k)P(k-1)x(k)}
\end{aligned}
\end{equation}
\begin{equation}
\begin{aligned}
\label{eqn30}
P(k)=\frac{\left[I-K(k)x'(k)\right]P(k-1)}{\gamma}
\end{aligned}
\end{equation}
Finally, using the process input $u(k)$ and process output $y(k)$, the numerator and denominator coefficients of the transfer function \eqref{eqn11} can be computed using RLS identification \cite{ref25}.
\subsection{Calculating Gains of the Controller}
\begin{figure}[h]
\centering
\includegraphics[trim=0.9cm 0.8cm 0.2cm 0.3cm, clip=true, totalheight=0.2\textheight, angle=0]{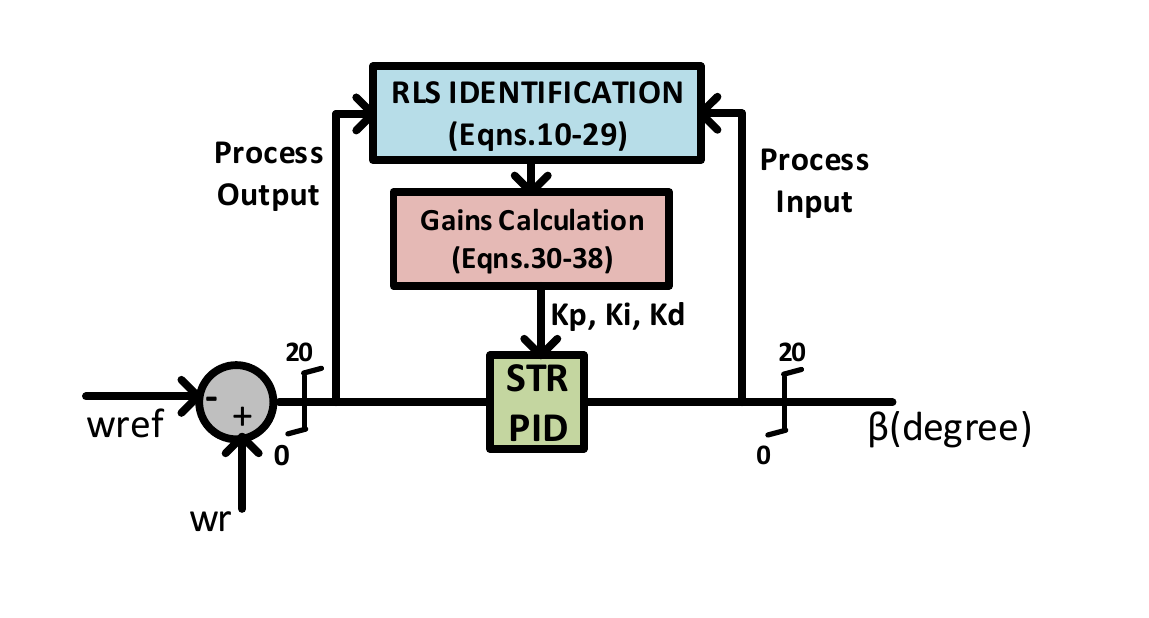}
\caption{Proposed STR Controller.}
\label{fig4}
\end{figure}

For calculating the gains of the controller, the process model is always restricted to second order. This algorithm calculates the proportional, integral, and derivative gains $K_p$, $K_i$, and $K_d$ every sample period. In this process, the closed loop pole shifting factor $\alpha$  is the only adjustment or tuning that is required. Using \eqref{eqn11} $2^{nd}$ order model can be represented as 
\begin{equation}
\begin{aligned}
\label{eqn31}
\frac{y}{u}=\frac{b_1q^{-1}+b_2q^{-2}}{1+a_1q^{-1}+a_2q^{-2}}=\frac{B}{A}  \text{(open loop)}
\end{aligned}
\end{equation}
From \eqref{eqn28}, the open loop characteristic equation is given by
\begin{equation}
\begin{aligned}
\label{eqn32}
1+a_1q^{-1}+a_2q^{-2}=0
\end{aligned}
\end{equation}
Thus, the closed loop characteristic equation using pole shifting by a factor $\alpha$ can be represented as
\begin{equation}
\begin{aligned}
\label{eqn33}
(1+\alpha q^{-1})(1+a_1\alpha q^{-1}+a_2\alpha^2 q^{-2})=0
\end{aligned}
\end{equation}
where, $0\le\alpha\le1$ and $q$ is a shift operator.
From the above, the control structure is given by
\begin{equation}
\begin{aligned}
\label{eqn34}
u(k)=\frac{T(q^{-1})}{R(q^{-1})}y_r(k)-\frac{S(q^{-1})}{R(q^{-1})}y(k)
\end{aligned}
\end{equation}
If in \eqref{eqn34} we let
\begin{equation}
\begin{aligned}
\label{eqn35}
R(q^{-1})=(1-q^{-1})(1+r_1q^{-1})
\end{aligned}
\end{equation}
\begin{equation}
\begin{aligned}
\label{eqn36}
S(q^{-1})=s_0+s_1q^{-1}+s_2q^{-2}
\end{aligned}
\end{equation}
where
$$s_0=T_sK_i+\frac{K_d}{T_s}+K_p$$
$$s_1=-\frac{2K_d}{T_s}-K_p+r_1K_p$$
$$s_2=\frac{K_d}{T_s}-r_1K_p$$
The architecture can be represented in terms of PID gains, which can be calculated using the following set of equalities:
\begin{equation}
\begin{aligned}
\label{eqn37}
K_i=\frac{-(s_0+s_1+s_2)}{T_s}
\end{aligned}
\end{equation}
\begin{equation}
\begin{aligned}
\label{eqn38}
K_p=\frac{(s_1+2s_2)}{1+r_1}
\end{aligned}
\end{equation}
\begin{equation}
\begin{aligned}
\label{eqn39}
K_p=T_s\left[\frac{r_1s_1-(1-r_1)s_2}{1+r_1}\right]
\end{aligned}
\end{equation}
As the system operating conditions changes, the coefficients of the transfer function get updated and hence the STR PID controller auto-tunes in real-time. The derivative part in PID controller helps in reducing the overshoot. Fig. \ref{fig4} shows the architecture of the proposed controller.
\section{Implementation of the Proposed Controller}
\begin{figure}[h]
\centering
\includegraphics[trim=0cm 0.6cm 0cm 0.6cm, clip=true, totalheight=0.25\textheight, angle=0]{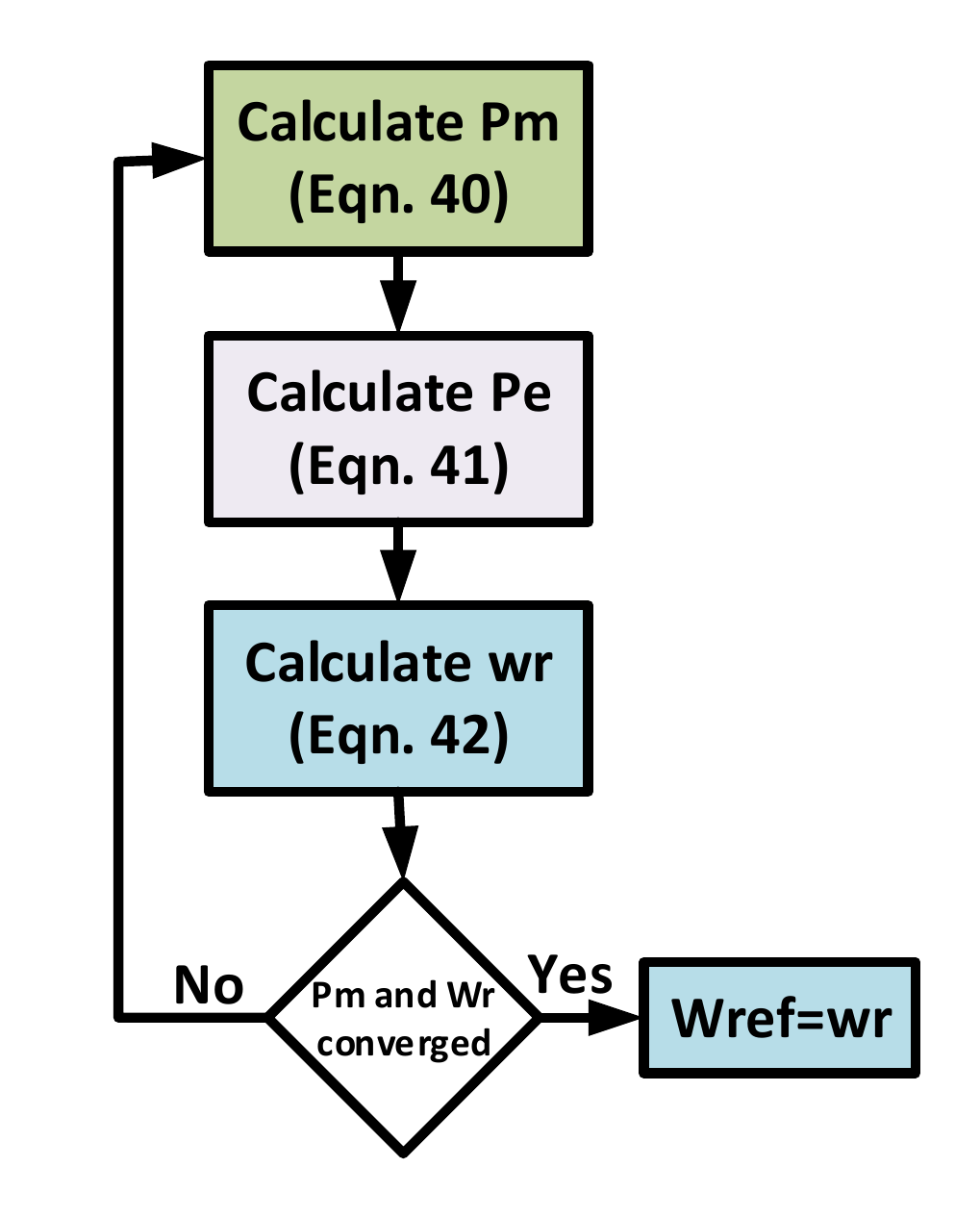}
\caption{Flowchart for $\omega_{ref}$ calculation.}
\label{fig6x}
\end{figure}

\begin{figure}[h]
\centering
\includegraphics[trim=0cm 0.6cm 0cm 0.6cm, clip=true, totalheight=0.25\textheight, angle=0]{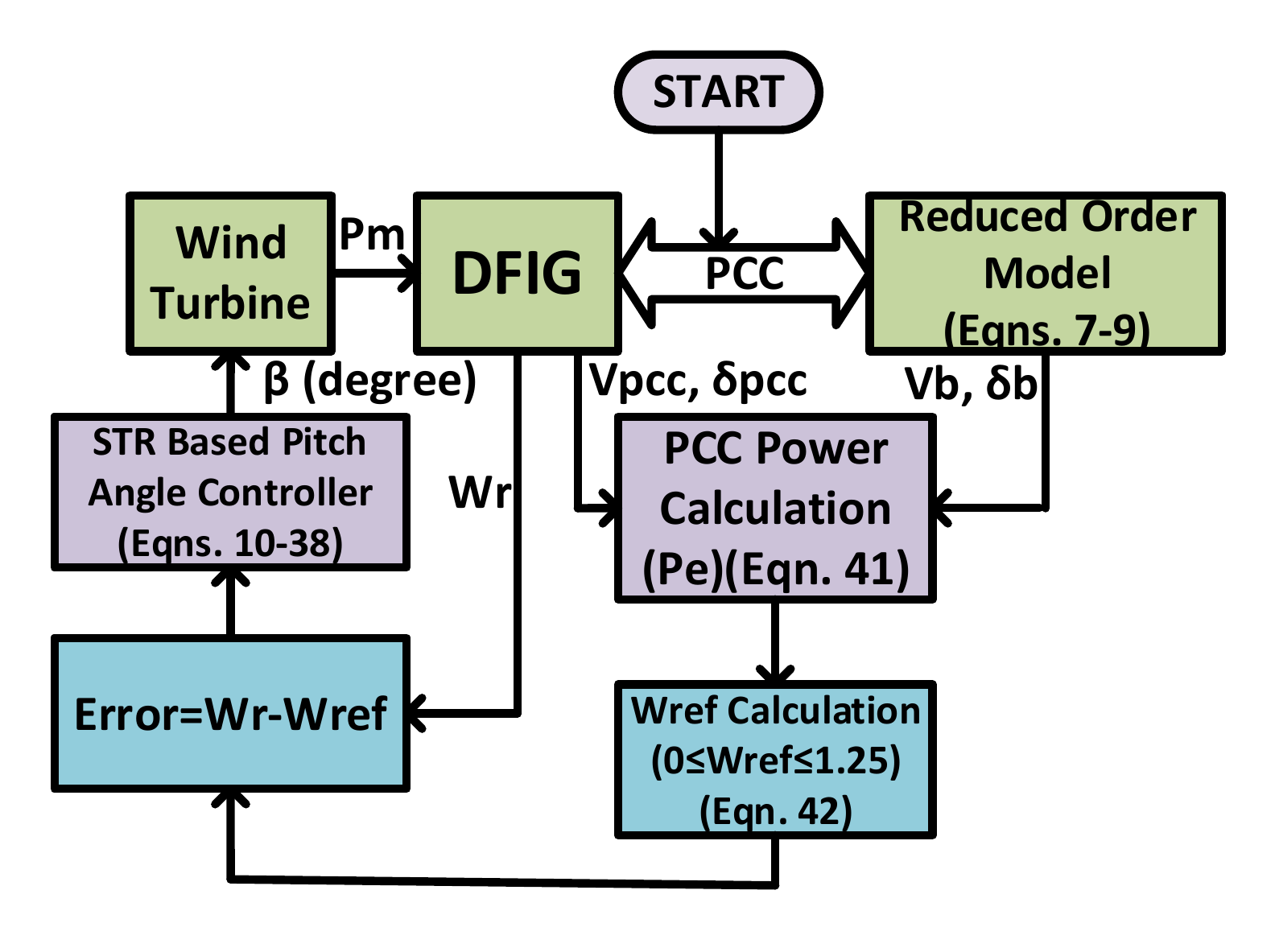}
\caption{Flowchart of the proposed controller.}
\label{fig6}
\end{figure}

The proposed adaptive pitch angle controller uses the difference between the reference speed $(\omega_{ref})$ and the actual speed ($\omega_r$) for estimating the control signal. The $\omega_{ref}$ is calculated as follows:\\
\textbf{Step:1} Initialize $\omega_r$ and estimate $P_m$ from \eqref{eqn4} and represented as:  
\begin{equation}
\label{eqn4x1}
P_m(t) = k_pc_p\left(\frac{\omega_r(t)}{r_{gear}\lambda}\right)^3
\end{equation}
At MPPT, $\lambda$, $c_p=1$ [p.u] and using \eqref{eqn4x1}, the mechanical power is represented as
\begin{equation}
\label{eqn6x1}
P_m(t) = k_p\left(\frac{\omega_r(t)}{r_{gear}}\right)^3
\end{equation}
where $t$ is the current iteration.\\
\textbf{Step:2} The electrical power delivered $P_e$ is calculated using grid conditions at boundary and PCC bus to include the grid transient effects in the controller action, whereas conventional pitch controller doesn't account for this calculation.  
\begin{equation}
\begin{aligned}
\label{eqn39}
P_e(t)=\frac{V_{pcc}(t)V_B(t)}{X}sin(\delta_B(t)-\delta_{pcc}(t))
\end{aligned}
\end{equation}
where $V_{pcc}$ and $V_B$  is the voltage of the WTG bus and boundary bus respectively,  $\delta_{pcc}$ and $\delta_B$   are the voltage angle at PCC and boundary bus respectively, and  $X$ is the reactance between PCC and boundary bus. Generally, stator resistance is small enough to ignore power loss associated with it and when the converter power loss is neglected, the total real power (here $P_e$ ) injected into the grid equals to the sum of the stator power and the rotor power \cite{ref26,ref27}.\\
\textbf{Step:3} Using the $P_m$ in \eqref{eqn6x1} and $P_e$ in \eqref{eqn39}, the $\omega_{r}$ is calculated as follows:
\begin{equation}
\label{eqn6n}
\omega_{r}(t+1)=\frac{P_m(t)-P_e(t)}{J\left(\frac{\omega_r(t)-\omega_r(t-1)}{\Delta t}\right)}
\end{equation}
where $J$ is the moment of inertia, $\Delta t$ is the simulation time step. Steps 1, 2 and 3 are repeated until $P_m$ and $\omega_r$ is converged and the converged value of $\omega_r$ is taken as the $\omega_{ref}$ (Fig. \ref{fig6x}). The integrated implementation flowchart is as shown in Fig. \ref{fig6}.

\section{Experimental Test Bed and Results}
\begin{figure}[h]
\centering
\includegraphics[width=3.5in,height=2.1in]{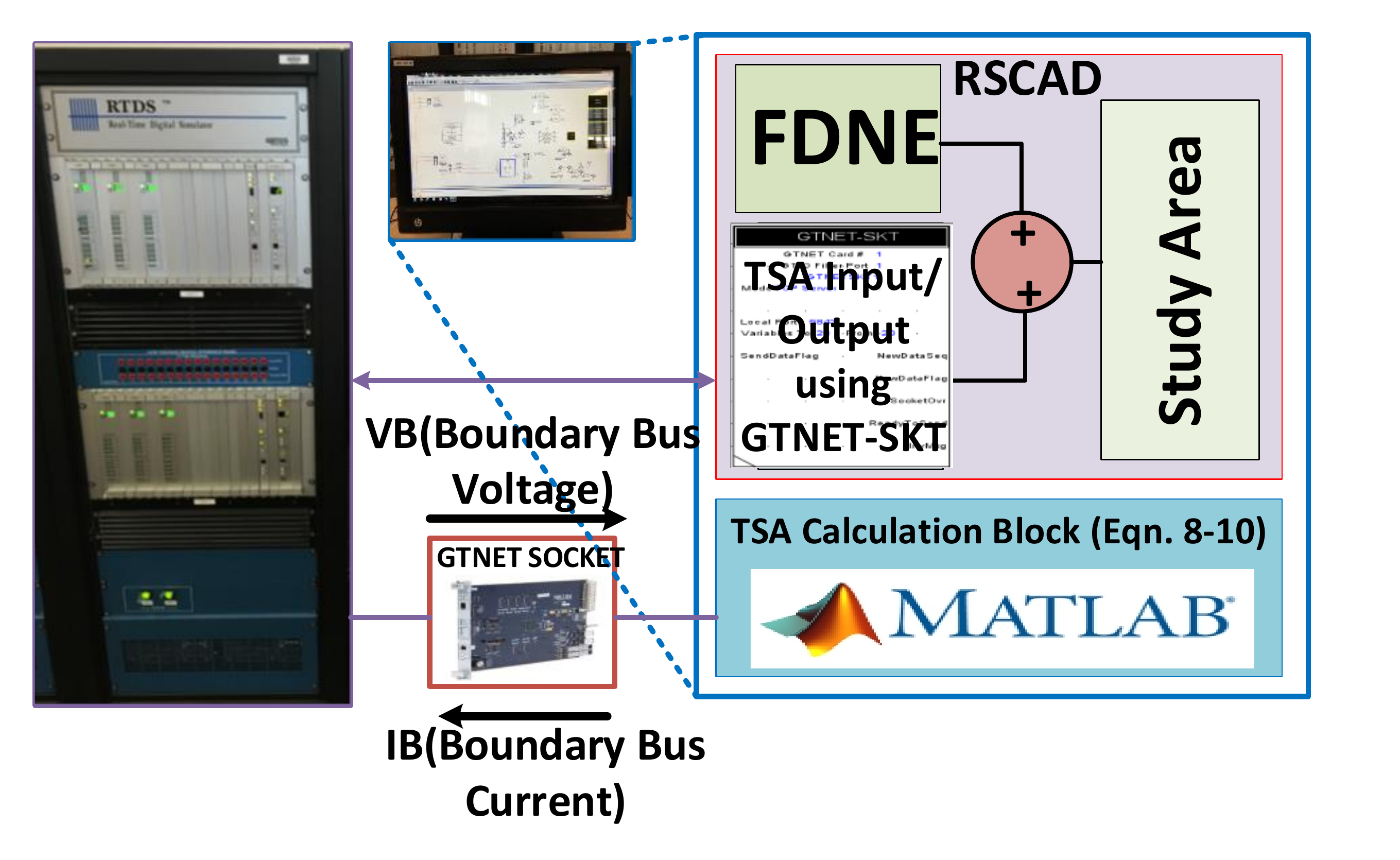}
\caption{Experiment setup in RTDS.}
\label{fig7}
\end{figure}

The proposed framework in Fig. \ref{fig6} is using a lab real-time simulator set up on Kundur's two-area \cite{ref28} and IEEE 39-bus \cite{ref29} test system models with WTGs. Table \ref{table1} and \ref{table2} show the simulation parameters of the wind turbine and DFIG. The real-time test bed consists of a) Reduced order RTDS/RSCAD\textsuperscript{TM} models of Kundur two area and IEEE 39-bus test systems, b) RTDS/RSCAD\textsuperscript{TM} model of WTGs and, c) GTNET-SKT connection between RTDS and MATLAB for interfacing TSA type equivalent with EMT type simulation (Fig. \ref{fig7}). The grid models are an actual representation of the wind farms and characterize real-time closed-loop control with real-life verified generator and control models with GE controllers. The operating principle of the test-bed is the rules that guide the machine model to work based on the grid changes.
\begin{table}[h]
\processtable{Simulation Parameters of Wind Turbine)\label{table1}}
{\begin{tabular*}{20pc}{@{\extracolsep{\fill}}ll@{}}\toprule
Parameter Name & Value\\
\midrule
Rated generator power & 2.2 MVA\\
\hline
Rated turbine power & 2.0 MW\\
\hline
Generator speed at rated turbine speed (p.u)  & 1.2 p.u.\\
\hline
Rated wind speed & 12.0 m/s\\
\hline
Cut-in wind speed & 6.0 m/s\\
\hline
Cut-out wind speed & 25 m/s\\
\hline
Rate of change of pitch angle & $\pm 10^0/s$\\
\botrule
\end{tabular*}}{}
\end{table}

\begin{table}[h]
\processtable{Simulation Parameters of DFIG)\label{table2}}
{\begin{tabular*}{20pc}{@{\extracolsep{\fill}}ll@{}}\toprule
Parameter Name & Value\\
\midrule
Rated stator voltage (L-L RMS) & 0.69 kV\\ 
\hline
Turn ratio (rotor over stator) & 2.6377\\
\hline
Rated MVA & 2.2 MVA\\
\hline
Stator resistance & 0.00462 p.u.\\ 
\hline
Stator leakage reactance & 0.102 p.u.\\ 
\hline
Unsaturated magnetizing reactance & 4.348 p.u.\\  
\hline
First cage rotor resistance & 0.006 p.u.\\
\hline
First cage rotor leakage reactance & 0.08596 p.u.\\ 
\hline
Inertia constant & 1.5 MWs/MVA\\
\botrule
\end{tabular*}}{}
\end{table}

\begin{figure}[h]
\centering
\includegraphics[width=3.5in,height=2.3in]{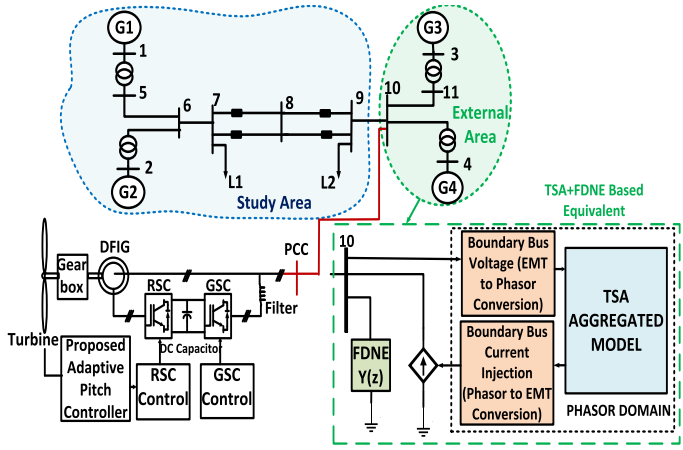}
\caption{Proposed dynamic equivalent of two-area test system.}
\label{fig5}
\end{figure}
\begin{figure}[h]
\centering     
\subfigure[Bus 7 RMS Voltage.]{\label{fig7a}\includegraphics[width=0.23\textwidth,height=1.55in]{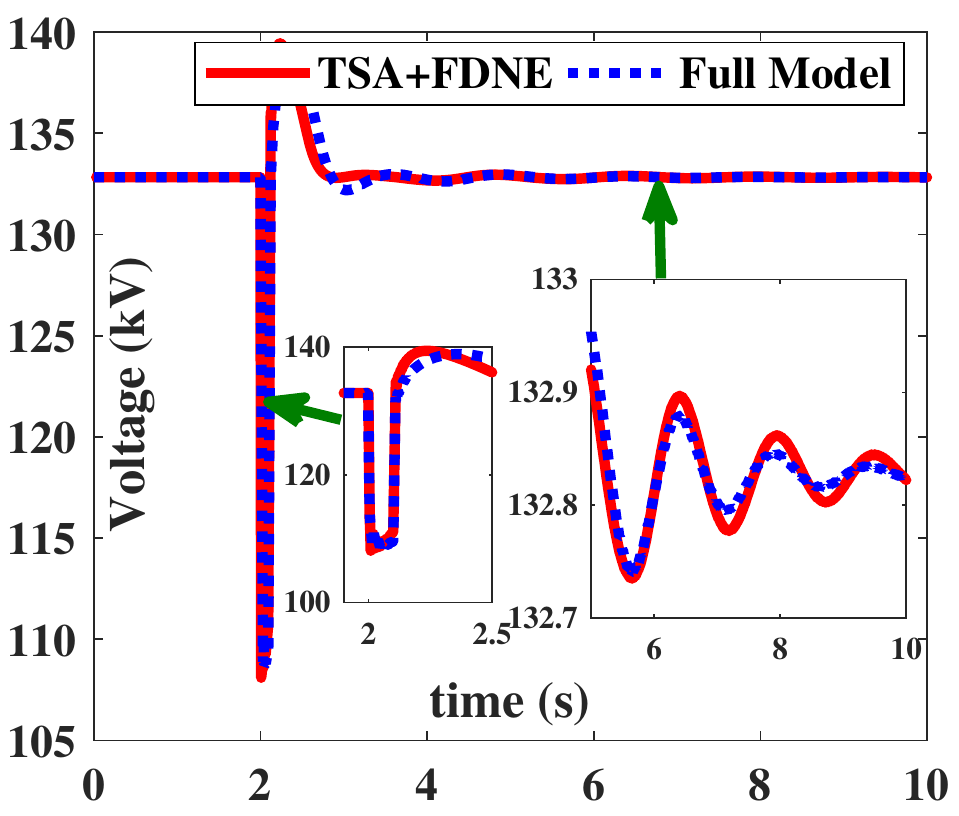}}
\subfigure[Relative Speed Generator-3 w.r.t Generator-2.]{\label{fig7b}\includegraphics[width=0.23\textwidth,height=1.55in]{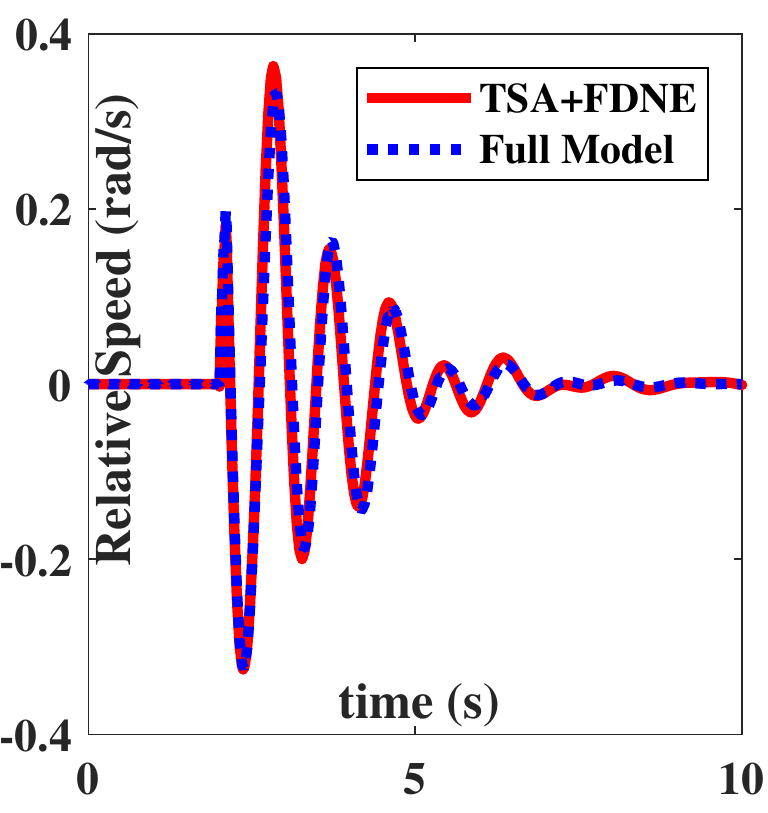}}
\caption{FDNE validation}
\end{figure}
\begin{figure}[h]
\centering     
\subfigure[Realistic wind speed pattern]{\label{fig11a}\includegraphics[width=0.23\textwidth,height=1.55in]{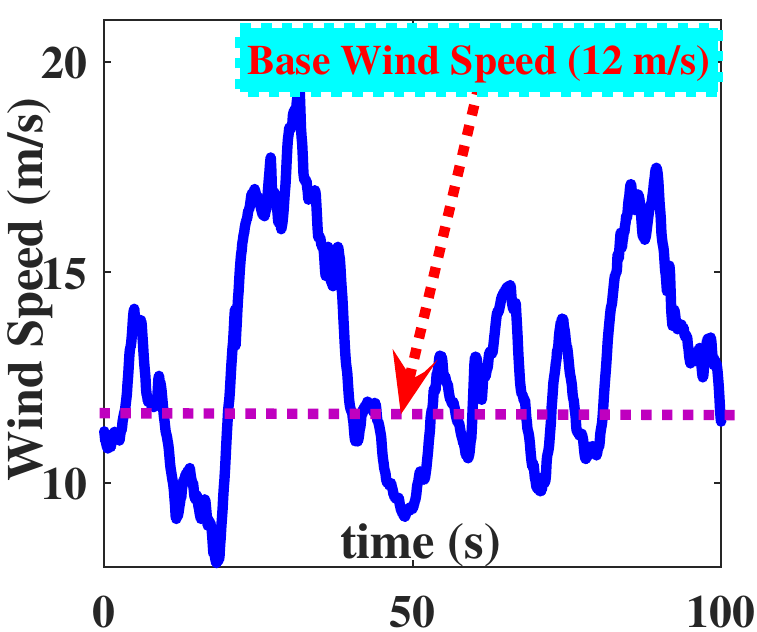}}
\subfigure[Active Power]{\label{fig11b}\includegraphics[width=0.23\textwidth,height=1.55in]{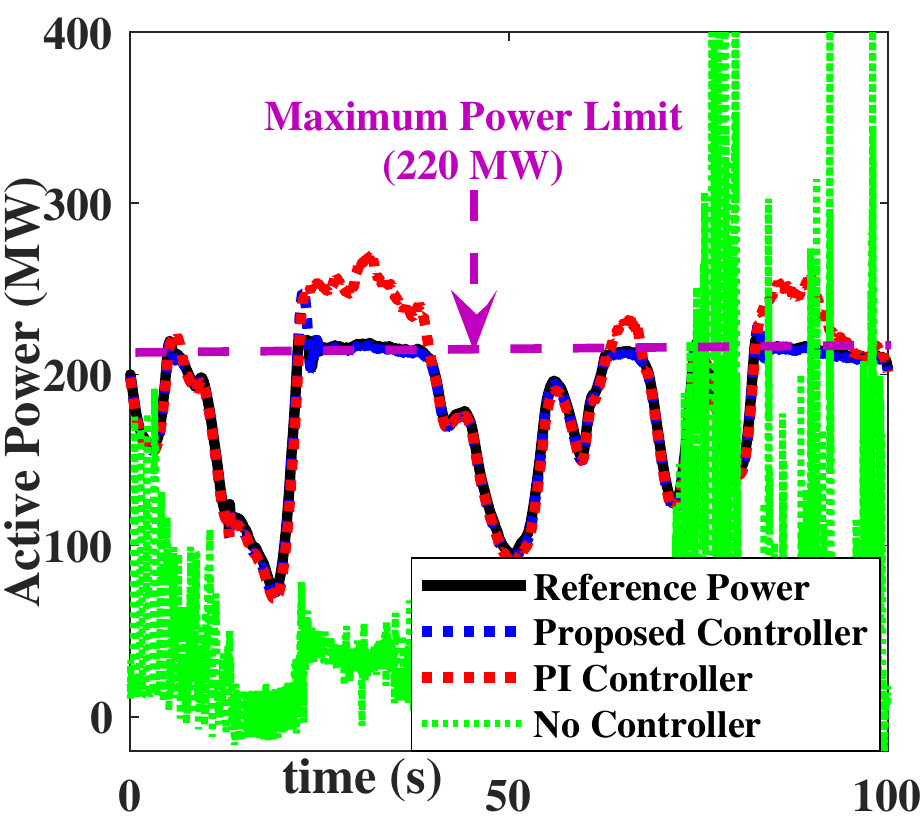}}
\caption{Wind Speed and DFIG Active Power}
\end{figure}
\subsection{Validation of the algorithm using Kundur's test system}
First, for validating the algorithm using grid integrated WTGs, two-area test system (see Fig. \ref{fig5}) is used. The test system consists of four 900MVA synchronous generators and a WTG at bus-10. Based on the location of WTG, the test system is divided into study and external area as shown in Fig. \ref{fig5}. The external area is modeled as a wide-band equivalent, which is the combination of TSA and FDNE. The TSA type equivalent is modeled in MATLAB\textsuperscript{\textregistered} in phasor domain and FDNE type equivalent is modeled in RSCAD\textsuperscript{TM} in EMT domain.
The reduced order model of the test system is validated by comparing its behavior under transient response with the original test system. For this, 3-phase bolted faulted is created at 0.1 sec for the duration of 0.1 sec. Fig. \ref{fig7a} and Fig. \ref{fig7b} shows the comparison of RMS voltage at bus 7 and the relative speed of Gen-3 w.r.t Gen-2 respectively. From the above results, it can be observed that the reduced order model behaves similarly as the full model under transient condition. Several other event analyses have been studied and similar results are obtained. 

\begin{figure}
\centering     
\subfigure[DFIG rotor speed]{\label{fig18a}\includegraphics[width=0.23\textwidth,height=1.55in]{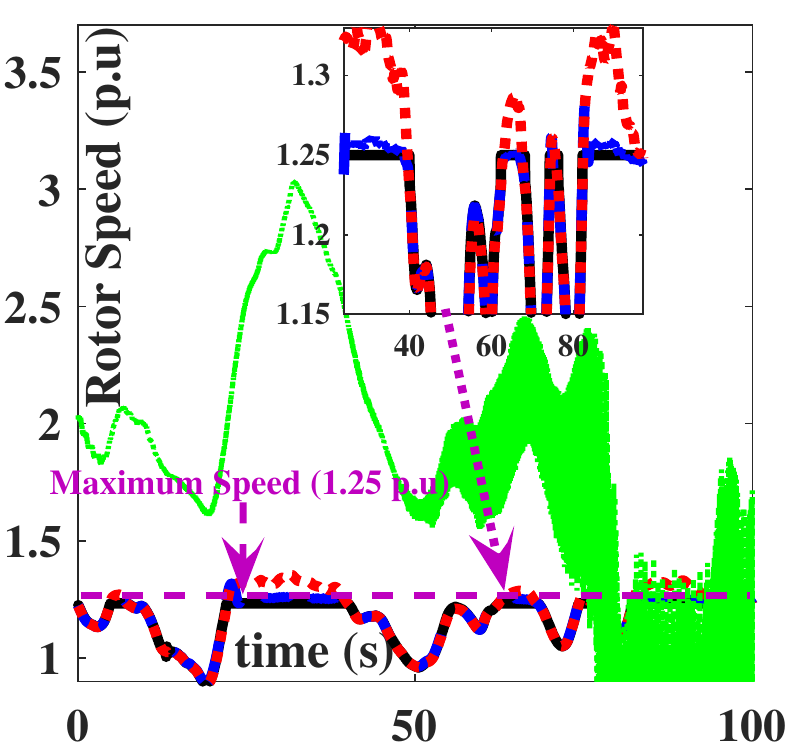}}
\subfigure[Relative speed of generator-1 w.r.t synchronous speed]{\label{fig18b}\includegraphics[width=0.23\textwidth,height=1.55in]{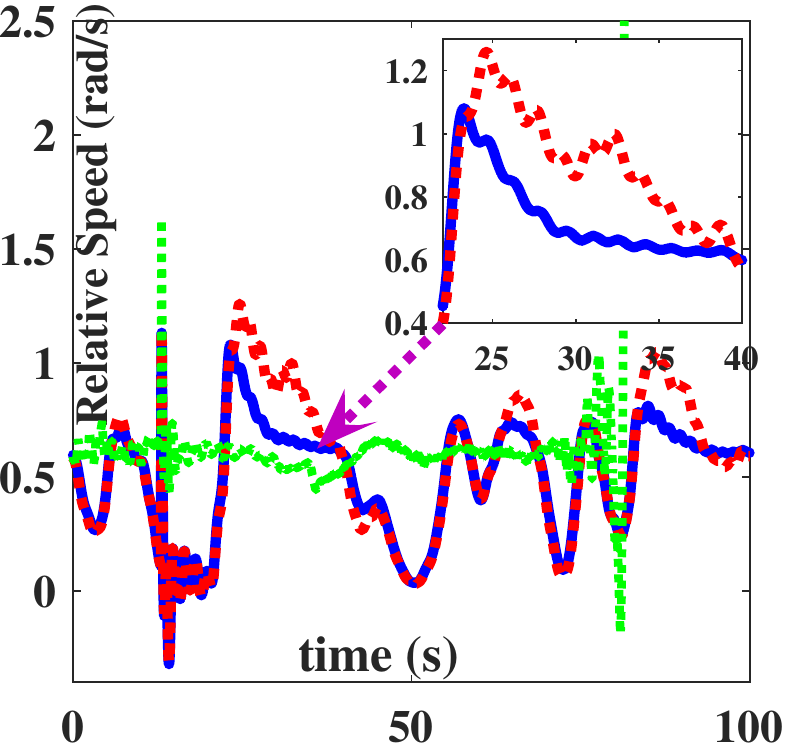}}
\subfigure{\includegraphics[width=3.3in,height=0.15in]{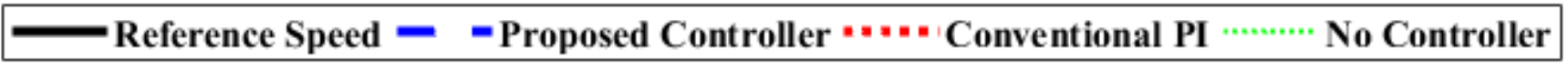}}
\caption{DFIG rotor speed and relative speed comparison.}
\end{figure}

\begin{figure}[!t]
\centering     
\subfigure[Pitch angle.]{\label{fig22a}\includegraphics[width=0.23\textwidth,height=1.55in]{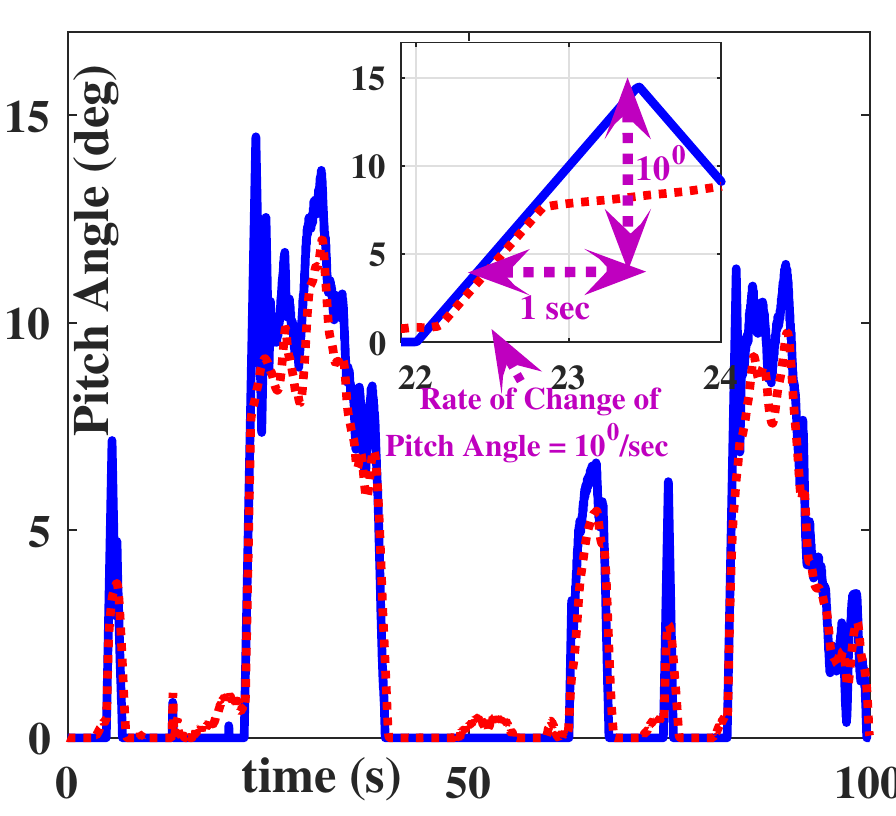}}
\subfigure[Mechanical torque of wind turbine.]{\label{fig22b}\includegraphics[width=0.23\textwidth,height=1.55in]{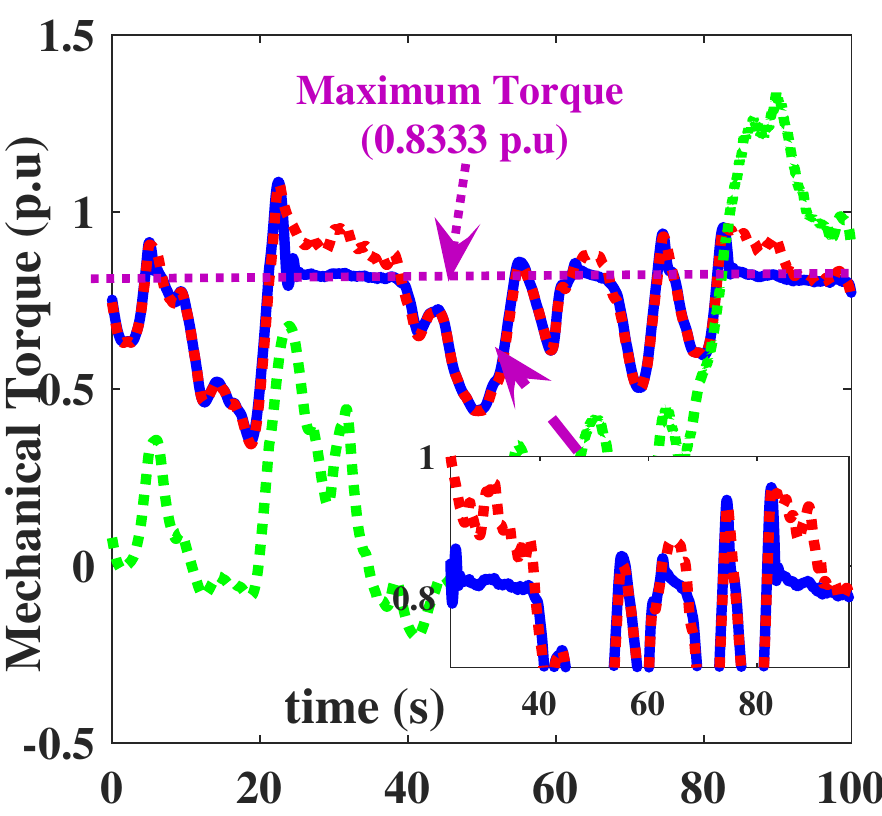}}
\subfigure{\includegraphics[width=3.3in,height=0.15in]{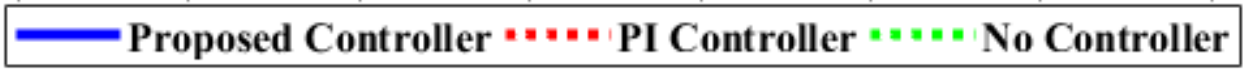}}
\caption{Pitch angle and mechanical torque comparison.}
\end{figure}

To validate the controller under rapidly varying realistic wind conditions Fig. \ref{fig11a} has been extracted from the ERCOT data along with a 3-phase bolted fault on Bus-8 for a duration of 6 cycles at 13 sec, and the performance is compared with conventional PI and no controller systems. Figs. \ref{fig11b}-\ref{fig18a} show the active power and rotor speed of the DFIG. It shows that with conventional controller the rating of the DFIG exceeds its limit and effectively increases stress on all connected electrical equipment. For example, the active power at 30sec with a proposed controller is 219.6 MW, whereas with a conventional controller it is 264.06 MW. So, with the conventional controller, the active power is 20\% more than the rated value which increases the stress on electrical equipment. Even the rotor speed crosses its limit when controlled by conventional PI controller (For example, it crosses 1.35 p.u at 30 sec while the limit is 1.25 p.u). So it can be concluded that the proposed controller controls the output power and at the same time limits the rotor speed. Additionally, other conventional generators (for example $G_1$ here) connected to the grid has less rotor oscillations with proposed controller (See Fig. \ref{fig18b}). Also, it can be seen from Fig. \ref{fig22a} that the rate of change of pitch angle is within its limit (10 deg/s). Additionally, Fig. \ref{fig22b} illustrates that the proposed controller is effectively limiting the mechanical torque. It can be observed that at 90 sec the mechanical torque with a conventional controller is 0.9266 p.u. Hence, the conventional controller provides fatigue caused by increased mechanical stress on the turbine due to torque overrun by 11.20\%. 
The RLS identification is performed for $\omega_r-\omega_{ref}$ and $\beta$  as shown in Fig. \ref{fig4}. The controller gains $K_i$, $K_p$, and $K_d$ are calculated at every time step using online identification routine. The conventional PI controller gains are adapted from GE wind turbine field implemented values \cite{ref4}.

\begin{figure}[!t]
\centering
\includegraphics[width=3.5in,height=2.0in]{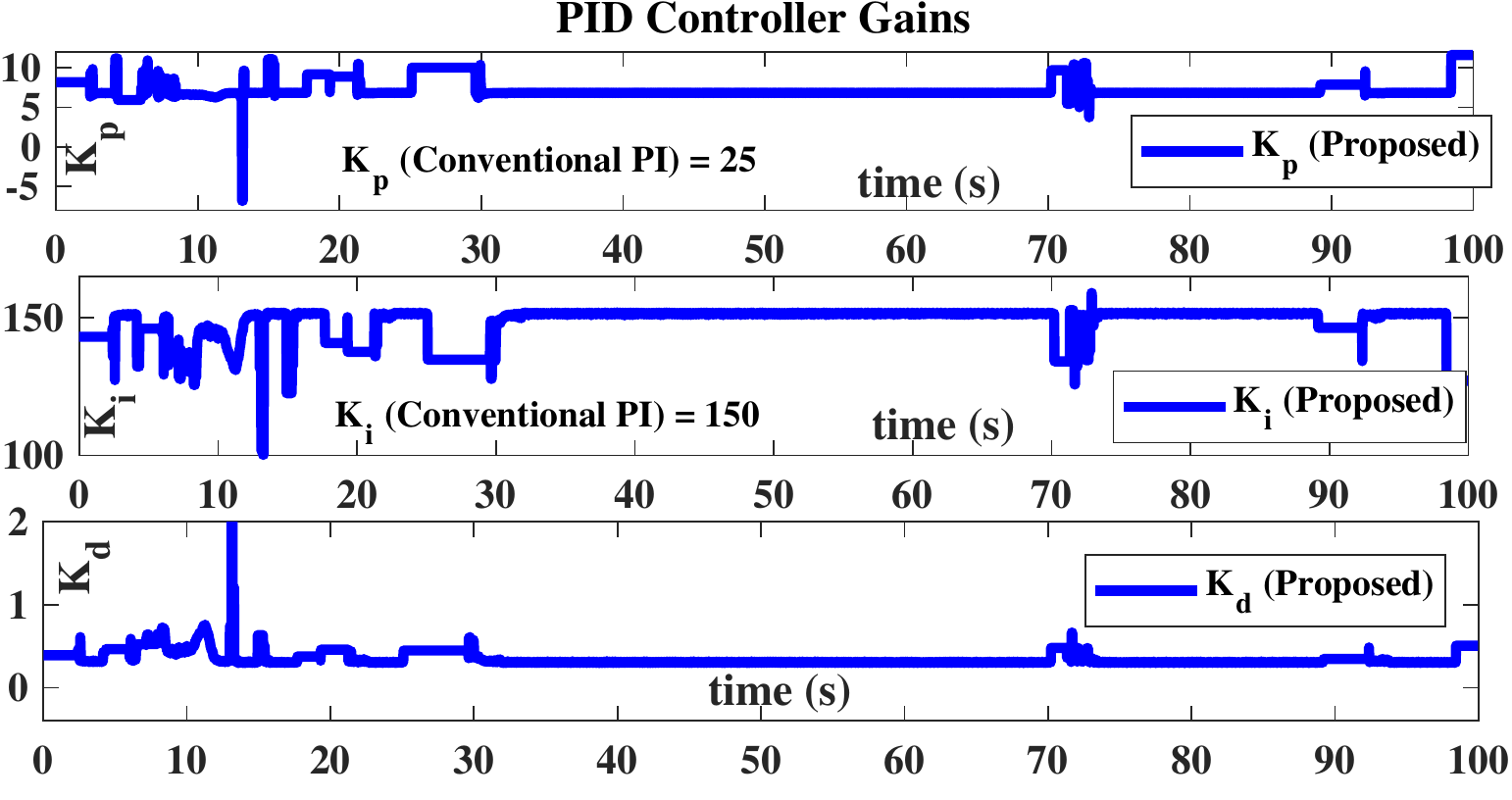}
\caption{STR and conventional PI controller gains.}
\label{fig10}
\end{figure}
Fig. \ref{fig10} shows the comparison of gains of STR and conventional PI controller for two-area system. From Fig. \ref{fig10}, it can be seen that STR controller auto-tunes as the operating condition changes whereas conventional PI controller has fixed gains irrespective of operating condition.  For reliable operation, the generator should be operated below the maximum speed limit (1.25 p.u) and thus tuning is necessary.
\subsection{Validation of the algorithm with IEEE 39-bus test system}
\begin{figure}[!t]
\centering
\includegraphics[width=3.5in,height=3in]{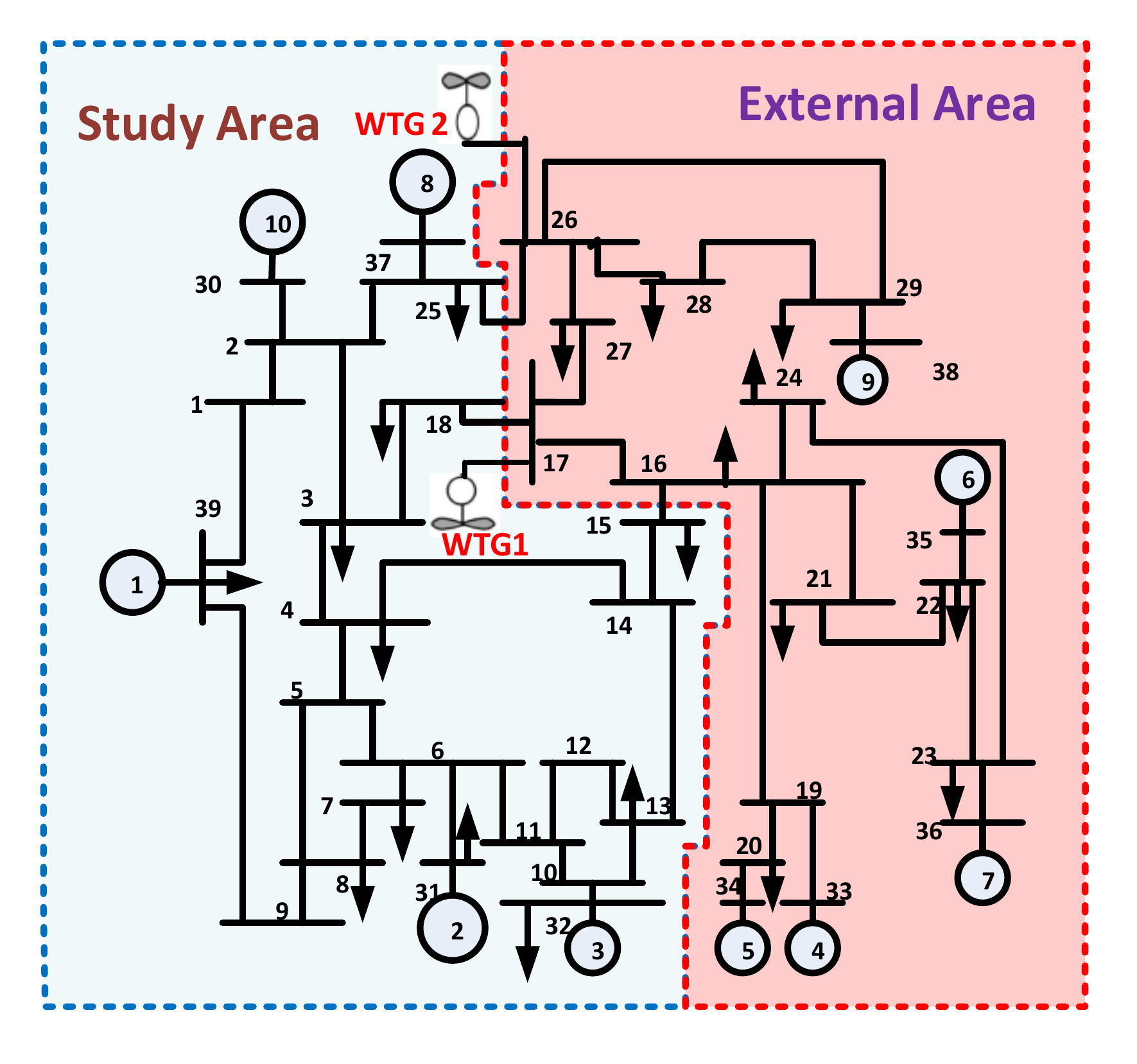}
\caption{IEEE 39-Bus system.}
\label{fig24}
\end{figure}

For further validation, the algorithm is also implemented on IEEE 39-bus system with WTGs connected at bus 17 and bus 26. The test system is divided into study and external area based on the location of the WTGs as shown in Fig. \ref{fig24}. The external area is modeled as a combination of TSA and FDNE.

\begin{figure}[!b]
\centering     
\subfigure[Realistic wind speed.]{\label{fig31a}\includegraphics[width=0.23\textwidth,height=1.55in]{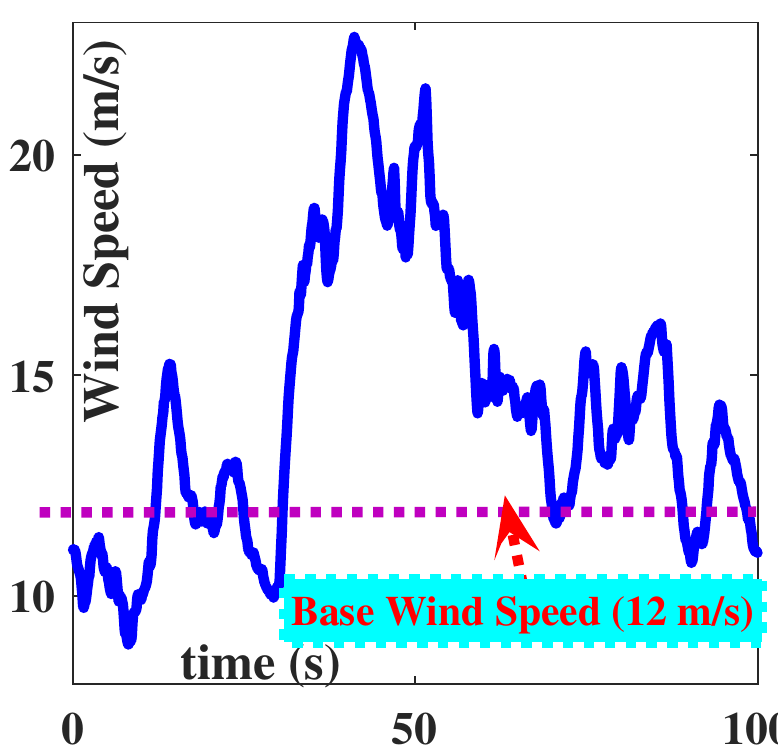}}
\subfigure[Bus-17 voltage (WTG-1).]{\label{fig31b}\includegraphics[width=0.23\textwidth,height=1.55in]{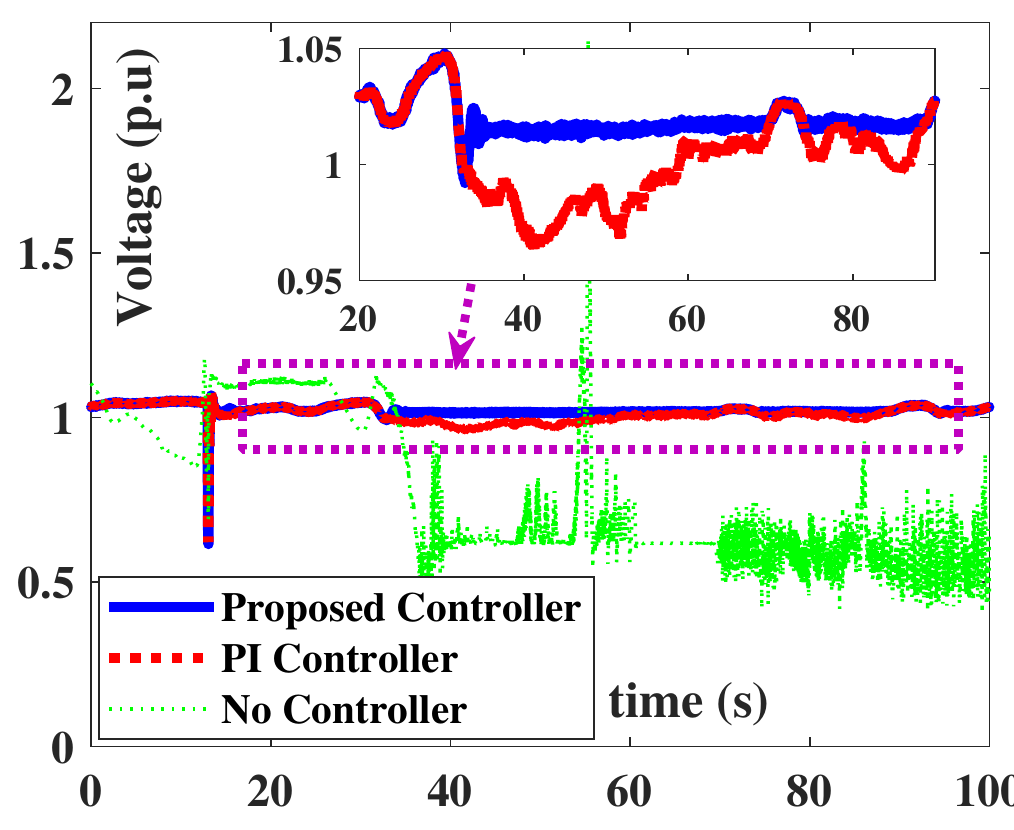}}
\caption{Wind speed and PCC voltage comparison.}
\end{figure}

\begin{figure}[!b]
\centering     
\subfigure[DFIG rotor speed (WTG-2).]{\label{fig33a}\includegraphics[width=0.23\textwidth,height=1.55in]{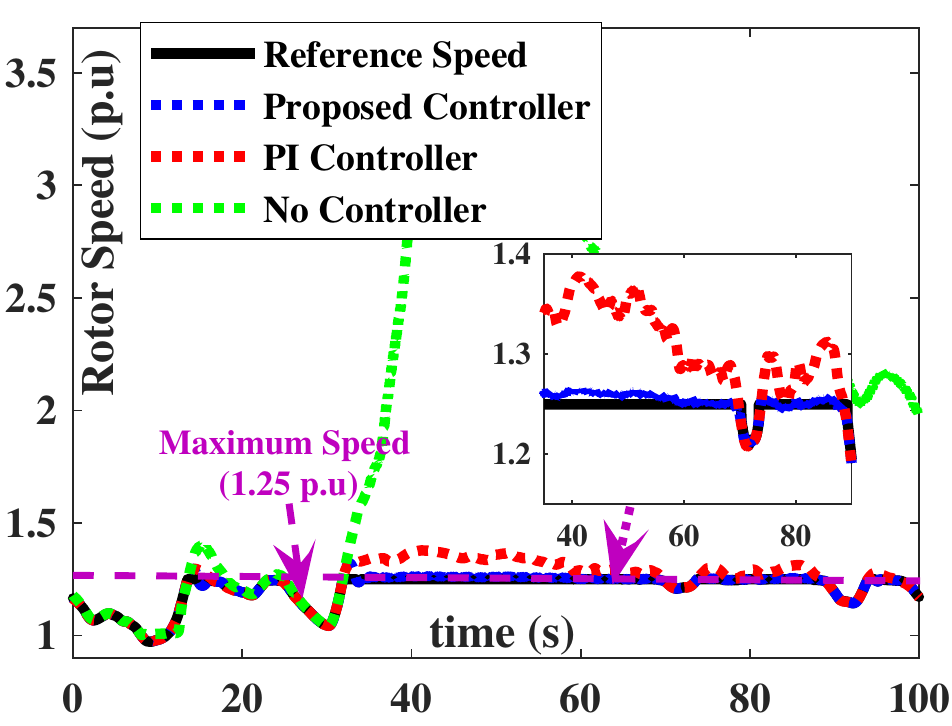}}
\subfigure[Active power (WTG-1).]{\label{fig33b}\includegraphics[width=0.23\textwidth,height=1.55in]{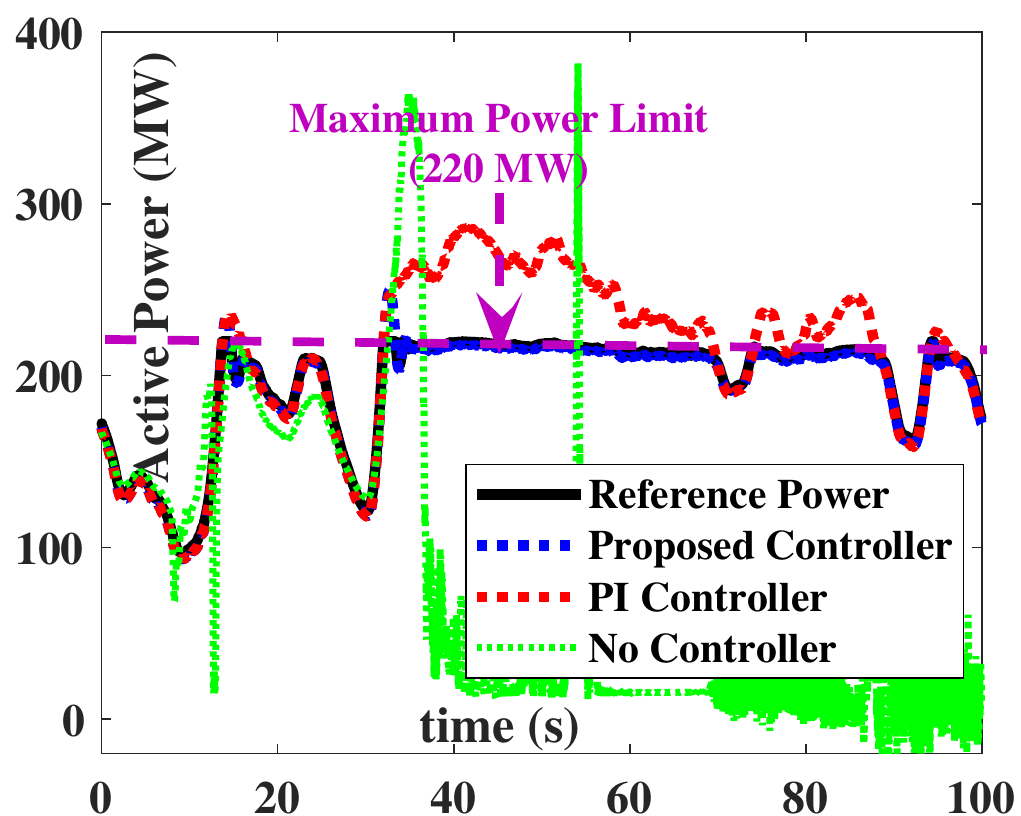}}
\caption{DFIG rotor speed and active power comparison.}
\end{figure}

\begin{figure}[!b]
\centering     
\subfigure[Relative speed of generator-3 w.r.t base speed (377 rad/s).]{\label{fig36a}\includegraphics[width=0.23\textwidth,height=1.55in]{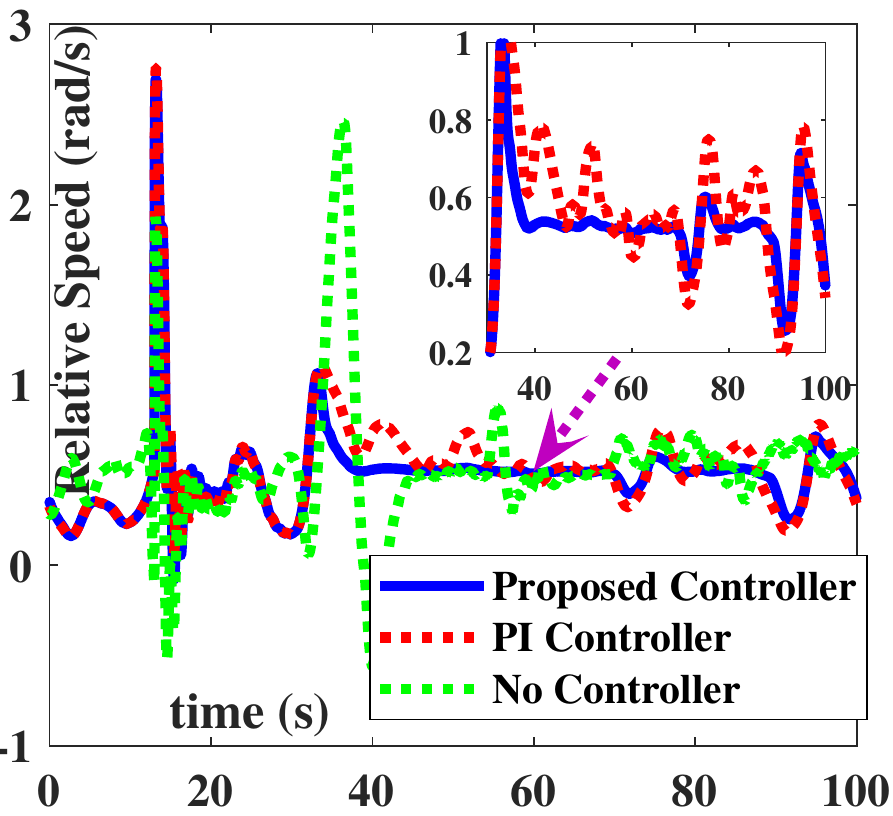}}
\subfigure[Mechanical torque of wind turbine (WTG-2).]{\label{fig36b}\includegraphics[width=0.23\textwidth,height=1.55in]{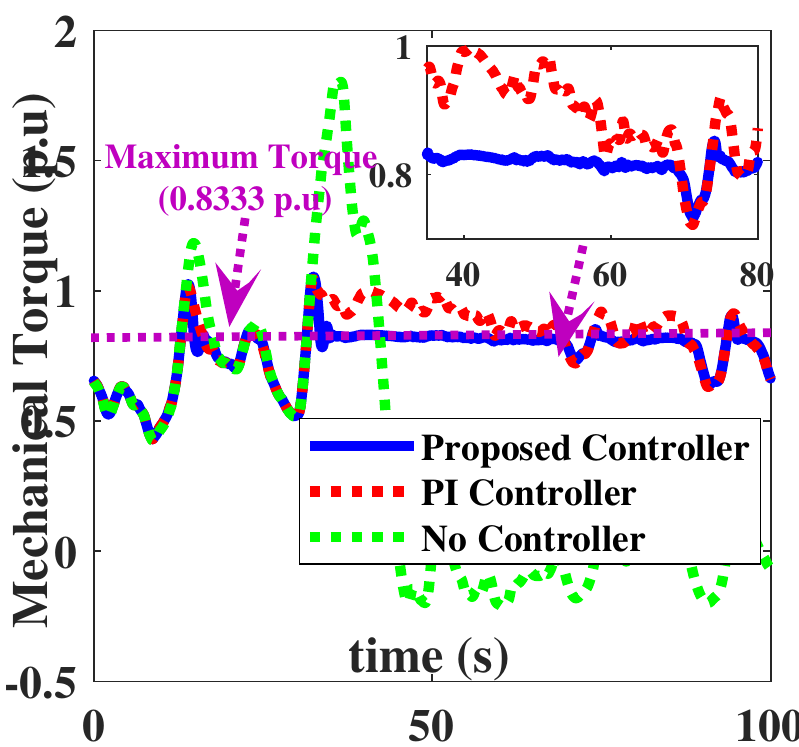}}
\caption{Relative speed and mechanical torque comparison.}
\end{figure}

In this case, the proposed controller is tested and validated for variable wind speed pattern (Fig. \ref{fig31a}) along with a 3-ph bolted fault on Bus-14 for 0.1 sec at 13sec. With the proposed controller, PCC voltage is much smoother and within allowable limit during high wind speed conditions when compared to PCC voltage with conventional PI controller (Fig. \ref{fig31b}). For  example,  the  voltage  at  40sec  with  the  proposed  controller  is  1.017 p.u, whereas  with  a  conventional  controller  it  is  0.9692 p.u.  So,with the conventional controller, the voltage is 6.266\% less than  the  steady state  value (1.034 p.u). Hence, the proposed controller improves the voltage by 4.93\% and can keep the voltage at the PCC within stable regions during high wind speed conditions. Fig. \ref{fig33a} shows the comparison of DFIG rotor speed of WTG-2 and Fig. \ref{fig33b} shows the comparison of the active power of WTG-1. Fig. \ref{fig36a} shows the relative speed of synchronous generator-3. Fig. \ref{fig36b} shows the mechanical torque of WTG-2. 

It can be seen that,  the  active power  at  45sec  with proposed  controller  is  218.7  MW, whereas  with  a  conventional  controller  it  is  285.8 MW.  So, with the conventional controller, the active power is 29.9\% more than  the  rated  value  which  increases  the  stress  on  electrical equipment.  Also, the  rotor  speed crosses its  limit using  conventional  controller  (for  example,  it  crosses  1.37p.u  at  45  sec where the limit is  1.25  p.u). The WTG control helps to keep the speed and active power under control and yet can keep the voltage at the PCC and other buses within stable operating region during high wind speed conditions. 
It can be observed that at 40 sec the mechanical torque with conventional controller is 0.9968 p.u, so the conventional controller provides fatigue caused by increased mechanical stress on the turbine due to torque overrun by 19.62\%. 

Hence, with the proposed controller, during high wind speed conditions, all electrical and mechanical parameters are within the rated limits so actions that are otherwise required to protect the electrical and mechanical equipment during these conditions is not a primary concern. Further with the proposed controller synchronous machine oscillations in the grid are damped out much faster when compared to conventional PI controller. The proposed architecture also ensures that the active power transfer is smooth thus maintaining the required power balance during high wind speed conditions.

\section{Conclusion}
The proposed sensor-less pitch angle control of WTG, considering the grid dynamics at wide band frequency and using STR controller is an efficient way of controlling the speed of the turbine during high wind speed. WTG is connected to reduced order model of power gird, the area in which WTG connected is modeled in detail while the remaining part is modeled as a combination of FDNE and coherency based TSA equivalent. The proposed method is validated in RTDS/RSCAD using WTG integrated reduced order models of Kundur two-area and IEEE-39 bus test systems. The results clearly illustrate that the proposed pitch angle controller provides better power balance, voltage regulation and reduces fatigue on the turbine. Additionally, the proposed architecture can work without anemometer, thus avoiding any kind of malfunctioning of the device. It has also been demonstrated that the architecture can be implemented in real-life as demonstrated using real-time simulators.

\bibliographystyle{iet}
\bibliography{Main.bib}

\begin{thebibliography}{10}

\bibitem{ref1}
`Electricity in the united states.'. (, .
\newblock Available from:
  \url{https://www.eia.gov/energyexplained/index.cfm?page=electricity_in_the_united_states}

\bibitem{ref2a}
Alhmoud, L.: `Reliability improvement for high-power igbt in wind energy
  applications', \emph{IEEE Trans Ind Electron},  2018, \textbf{PP}, pp.~1--9

\bibitem{ref2}
Zhang, J., Cheng, M., Chen, Z., Fu, X.
\newblock `Pitch angle control for variable speed wind turbines'.
\newblock In: Proc. 3rd Int. Conf. Elect. Utility DRPT. (Nanjing, China,  2008.
  pp.~ 2691--2696

\bibitem{ref4}
Ghosh, S., Kamalasadan, S., Senroy, N., Enslin, J.: `Doubly fed induction
  generator (dfig)-based wind farm control framework for primary frequency and
  inertial response application', \emph{IEEE Trans Power Syst},  2016,
  \textbf{31}, pp.~1861--1871

\bibitem{ref3}
Zhang, Y., Gole, A.M., Wu, W., Zhang, B., Sun, H.: `Development and analysis of
  the applicability of a hybrid transient simulation platform combining tsa and
  emt elements', \emph{IEEE Trans Power Syst},  2013, \textbf{28}, pp.~357--366

\bibitem{ref5}
Wang, S., Lu, S., Zhou, N., Lin, G., Elizondo, M., Pai, M.A.: `Dynamic-feature
  extraction, attribution, and reconstruction (dear) method for power system
  model reduction', \emph{IEEE Trans Power Syst},  2014, \textbf{29},
  pp.~2049--2059

\bibitem{ref6}
Liang, Y.F., Lin, X., Gole, A.M., Yu, M.: `Improved coherency based wide-band
  equivalents for real-time digital simulators', \emph{IEEE Trans Power Syst},
  2011, \textbf{26}, pp.~1410--1417

\bibitem{ref7}
Zhang, Y., Cheng, M., Chen, Z.: `Load mitigation of unbalanced wind turbines
  using pi-r individual pitch control', \emph{IET Renewable Power Generation},
  2015, \textbf{9}, pp.~262--271

\bibitem{ref8}
Van, T.L., Nguyen, T.H., Lee, D.C.: `Advanced pitch angle control based on
  fuzzy logic for variable-speed wind turbine systems', \emph{IEEE Trans Energy
  Convers},  2015, \textbf{30}, pp.~578--587

\bibitem{ref9}
Lasheen, A., Elshafei, A.L.: `Wind-turbine collective-pitch control via a fuzzy
  predictive algorithm', \emph{Renewable Energy},  2016, \textbf{87},
  pp.~298--306

\bibitem{ref10}
Ren, Y., Li, L., Brindley, J., Jiang, L.: `Nonlinear pi control for variable
  pitch wind turbine', \emph{Control Engineering Practice},  2016, \textbf{50},
  pp.~84--94

\bibitem{ref11}
Das, K.K., Buragohain, M.: `An algorithmic approach for maximum power point
  tracking of wind turbine using particle swarm optimization', \emph{IJAREEIE},
   2015, \textbf{4}, pp.~4099--4106

\bibitem{ref12}
Soued, S., Ebrahim, M.A., Ramadan, H.S., Becherif, M.: `Optimal blade pitch
  control for enhancing the dynamic performance of wind power plants via
  metaheuristic optimisers', \emph{IET Electric Power Application},  2017,
  \textbf{11}, pp.~1432--1440

\bibitem{ref13}
Thakallapelli, A., Ghosh, S., Kamalasadan, S.
\newblock `Real-time reduced order model based adaptive pitch controller for
  grid connected wind turbines'.
\newblock In: Proc. IEEE Industry Applications Society Annual Meeting.
  (Portland, USA,  2016. pp.~ 1--8

\bibitem{ref14}
Ghosh, S., Senroy, N.: `Electromechanical dynamics of controlled variable speed
  wind turbines', \emph{IEEE Syst Journal},  2015, \textbf{9}, pp.~639--646

\bibitem{ref16}
`Wind-turbine driven doubly-fed induction generator user manual'. (,  2015

\bibitem{ref17}
Slootweg, J.G., de~Haan, S.W.H., Polinder, H., Kling, W.L.: `General model for
  representing variable speed wind turbines in power systems dynamics
  simulations', \emph{IEEE Trans Power Syst},  2003, \textbf{18}, pp.~144--151

\bibitem{ref16a}
Chen, J., Lin, T., Wen, C., Song, Y.: `Design of a unified power controller for
  variable-speed fixed-pitch wind energy conversion system', \emph{IEEE Trans
  Ind Electron},  2016, \textbf{63}, pp.~4899--4908

\bibitem{ref16b}
Chen, J., Chen, J., Gong, C.: `New overall power control strategy for
  variable-speed fixed-pitch wind turbines within the whole wind velocity
  range', \emph{IEEE Trans Ind Electron},  2013, \textbf{60}, pp.~2652--2660

\bibitem{ref18}
`Wind turbine, documentation simpowersystems'. (,  2004

\bibitem{ref15}
Pena, R., Clare, J.C., Asher, G.M.
\newblock `Doubly fed induction generator using back-to-back pwm converters and
  its application to variable speed wind-energy generation'.
\newblock In: IEE Proceedings - Electric Power Applications. (,  1996. pp.~
  231--241

\bibitem{ref19}
Gole, A.. `Vector controlled doubly fed induction generator for wind
  applications'. (,  2004

\bibitem{ref20}
Woodford, D.A.. `Determination of main parameters for a doubly fed induction
  generator for a given turbine rating'. (,  2004

\bibitem{ref21}
Thakallapelli, A., Hossain, S.J., Kamalasadan, S.
\newblock `Coherency based online wide area control of renewable energy
  integrated power grid'.
\newblock In: Proc. IEEE PEDES. (Trivandrum, India,  2016. pp.~ 1--6

\bibitem{ref22}
{Chow, J. H. }: `Power System Coherency and Model Reduction'.
\newblock Power Electronics and Power Systems. ({Springer New York},  2014).
\newblock Available from: \url{https://books.google.com/books?id=HGnABAAAQBAJ}

\bibitem{ref23}
Thakallapelli, A., Ghosh, S., Kamalasadan, S.
\newblock `Real-time frequency based reduced order modeling of large power
  grid'.
\newblock In: Proc. Power and Energy Society General Meeting. (Boston, USA,
  2016. pp.~ 1--5

\bibitem{ref24}
Thakallapelli, A., Kamalasadan, S.
\newblock `Optimization based real-time frequency dependent reduced order
  modeling of power grid'.
\newblock In: Proc. Power and Energy Society General Meeting. (Chicago, USA,
  2017. pp.~ 1--5

\bibitem{ref25}
{K.  J.  Astrom and B.  Wittenmark}: `Adaptive control'.
\newblock ({Addison-Wesley Publishing Company},  1995)

\bibitem{ref26}
Lei, Y., Mullane, A., Lightbody, G., Yacamini, R.: `Modeling of the wind
  turbine with a doubly fed induction generator for grid integration studies',
  \emph{IEEE Trans Energy Convers},  2006, \textbf{21}, pp.~257--264

\bibitem{ref27}
Muller, S., Deicke, M., Doncker, R.W.D.: `Doubly fed induction generator
  systems for wind turbines', \emph{IEEE Ind Appl Mag},  2002, \textbf{8},
  pp.~26--33

\bibitem{ref28}
{P.  Kundur}: `Power System Stability and Control'.
\newblock ({New York: McGraw-Hill},  1994)

\bibitem{ref29}
Hiskens, I.. `Report: 39-bus system (new england reduced model)'. (,  2013

\end{thebibliography}
\vfill\pagebreak

\end{document}